\documentclass[%
 reprint,
 amsmath,amssymb,
 aps,
]{revtex4-2}

\usepackage{graphicx} 
\usepackage{dcolumn} 
\usepackage{bm} 
\usepackage{verbatim}
\usepackage{braket} 
\usepackage{enumitem} 
\usepackage{soul} 
\usepackage{textcomp, gensymb} 
\usepackage{xcolor} 
\usepackage{hyperref}

\usepackage[font=small]{caption}
\usepackage[labelformat=empty, position=top]{subcaption}
\usepackage[export]{adjustbox}

\begin{document}

\preprint{APS/123-QED}

\title{Investigation of bremsstrahlung emission in an electron cyclotron resonance ion source and its dependence on the magnetic confinement}

\author{Andrea Cernuschi}
 \email{andrea.cernuschi@lpsc.in2p3.fr}
 \altaffiliation[Also at ]{Department of Energy, Politecnico di Milano, Milan, Italy.}
\author{Thomas Thuillier}%
 \email{thomas.thuillier@lpsc.in2p3.fr}
\affiliation{%
 Université Grenoble Alpes, CNRS, Grenoble INP, LPSC-IN2P3, 38000 Grenoble, France
}%


\begin{abstract}
A Monte Carlo (MC) code is used to investigate the bremsstrahlung x-ray emission of an electron cyclotron resonance ion source (ECRIS) and its dependence on the axial magnetic confinement. The x-ray spectral temperature $T_S$ measured with the simulations is in fair agreement with previous experiments.
The dependence of $T_S$ on the minimum magnetic field of the configuration $B_{min}$ is corroborated, also observing that the ion extraction peak field $B_{ext}$ has no influence on temperature.
Details on the mechanism generating the hot electron population responsible for the change in spectral x-ray temperature as a function of $B_{min}$ are proposed, based on an in-depth investigation of the electron population with the MC code using a high statistics.
\end{abstract}

\maketitle



\section{Introduction}
\label{sec:introduction}

\begin{figure}[b]
\includegraphics[width=0.4\textwidth]{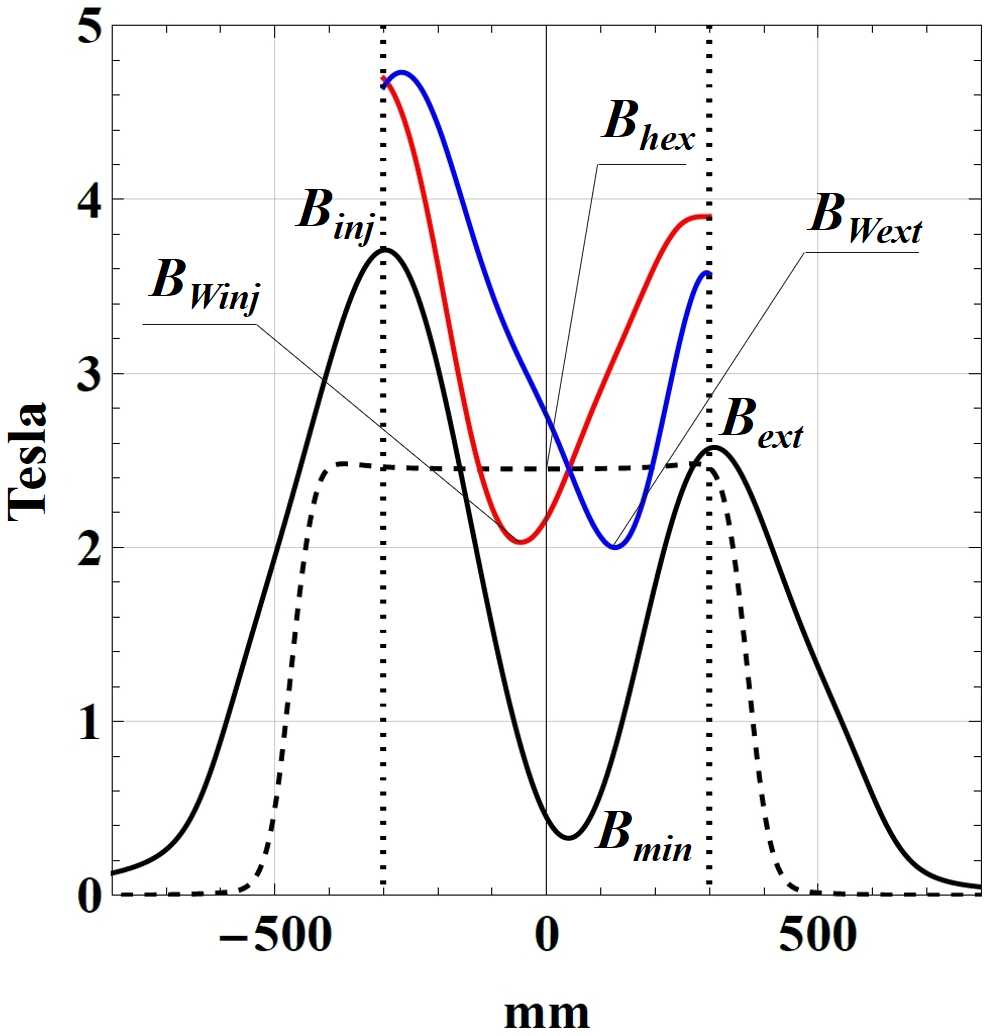}
\caption{\textbf{(Solid Black)} ECRIS typical axial magnetic field profile. The axial peak mirror field names $B_{inj}$, $B_{ext}$ and $B_{min}$ respectively stand for the atom injection and ion extraction field, and the central magnetic well minimum. \textbf{(Black Dashed)} Norma of the radial hexapolar magnetic field taken along a magnetic pole at the plasma chamber wall in the absence of axial magnetic field. The plateau hexapolar magnetic field value in the source is labelled $B_{hex}$. \textbf{(Solid red and blue)} Magnetic field norma taken at the radial wall along two consecutive hexapole poles in the presence of both the axial and the radial magnetic field. The B-minimum of the red and blue curves are respectively labelled $B_{Winj}$ and $B_{Wext}$. The \textbf{(Black dotted)} Lines display the axial injection \textbf{(left)} and extraction \textbf{(right)} limits of the ion source plasma chamber.}
\label{fig:B_asterics}
\end{figure}

Electron cyclotron resonance ion sources (ECRIS) are used to produce multicharged ion beams for particle accelerators~\cite{Geller_book}. The ions are extracted from a magnetized plasma, sustained by a constant gas feed and an injected microwave (with frequency $f_{RF}$), which heats the electrons through the ECR mechanism in a cylindrical cavity. The magnetic structure of ECRIS forms a minimum-B configuration, created by the superposition of an axial magnetic mirror and a radial hexapole, whose magnetic field is minimum at the center and maximum at the plasma chamber walls.
 Fig.~\ref{fig:B_asterics} presents the axial magnetic field intensity taken along the ion source axis (black solid line), usually generated by a set of 3 axial coils, with the middle one set to an opposite polarity to generate a minimum intensity in the center. The radial field intensity along the plasma chamber wall is plotted with a dashed black line; this profile is obtained when only the hexapole coil is energized. The profiles of the resulting total magnetic field norma along two consecutive hexapole poles at the wall are displayed in red and blue.

The ECRIS magnetic field is commonly described by 5 magnetic parameters: $B_{ecr}$, $B_{inj}$, $B_{min}$, $B_{ext}$ and $B_{hex}$, being respectively the magnetic field intensity associated with the ECR heating mechanism ($B_{ecr}=2\pi f_{RF} m_e /e$, with $m_e$ and $e$ being the electron mass and elementary charge), the peak axial magnetic field on the gas and microwave injection side, the minimum magnetic intensity at the center of the plasma, the axial peak intensity on the side of the ion extraction and the intensity of the hexapolar magnetic field at the plasma chamber wall. The minimum intensities of the red and blue curves are noted as $B_{Winj}$ and $B_{Wext}$ respectively. 
The axial positions associated with the injection, minimum and extraction peaks are named $z_{inj}$, $z_{min}$ and $z_{ext}$ respectively. The last closed magnetic surface in the cavity is $B_W~=~Min(B_{Winj},B_{Wext})$.
\\
The quality of the magnetic confinement is expressed as the minimum mirror ratio of the minimum-B structure $R=Min(B_W,B_{ext})/B_{min}$.
The relative intensity of the ECRIS magnetic parameters has been extensively studied, resulting in generic magnetic scaling laws which maximize the multicharged ion beam production in ECRIS~\cite{Hitz02}. Some standard guidelines for magnetic parameters in ECRIS are $B_W\approx 2 B_{ecr}$, $B_{hex}\approx 2 B_{ecr}$, $B_{inj} \approx 3-4 B_{ecr}$, $B_{ext}\gtrsim 2 B_{ecr}$ and $B_{min}\approx 0.4-0.8 B_{ecr}$.
\\
The ECR magnetic surface in ECRIS (defined as $B_{ecr}=const$) is closed and all the confined particles in the dense plasma pass through it. The averaged magnetic field direction seen by moving charged particles in ECRIS is axial. The ECR heating favors the build up of electron energy perpendicular to the magnetic field lines, leading to an anisotropic electron velocity distribution. The ECRIS plasma is inhomogeneous (due to the complicated minimum-B structure) and out of thermal equilibrium: it is composed of cold ions and a mixture of cold, warm and hot electrons. 
Modern ECRIS are operated at high microwave frequency ($f_{RF} \geq\,24\, GHz$) to obtain a dense plasma and thus enhance the multicharged ion beam intensity, since the latter is proportional to the square of the plasma density~\cite{Geller_book}. These new generation ECRIS are prone to a strong parasitic emission of bremsstrahlung radiation~\cite{Lyneis06}. The photons are either created by the volume interaction of the electron population with the ion species and the electrons of the plasma or by the final surface interaction of the electrons with the plasma chamber wall, when they are eventually deconfined out of the plasma.
\\
The ECRIS plasma is experimentally difficult to investigate, as the plasma chamber is surrounded by bulky magnets used to generate the minimum-B magnetic field. The line of sight toward the ECR plasma is usually reduced to the tiny plasma electrode circular hole from which the ion beams are extracted, which is nevertheless sufficient to perform insightful photon spectroscopy.
The bremsstrahlung spectral temperature of modern ECRIS has been measured as high as $T_S\approx$ 100 keV~\cite{Lyneis06}, with some photons detected above 1 MeV.
It has been next demonstrated that the photon spectral temperature in ECRIS is only a function of $B_{min}$~\cite{Benitez17} (see Fig.~\ref{fig:B_asterics}), rather than $B_{min}/B_{ecr}$ or the value of the axial magnetic field gradient taken on the ion source axis at the location of the ECR zone $\frac{dB_z}{dz}(z_{ECR})$.
A recent study has shown a linear dependence between the value of $B_{min}$ and the surface weighted mean magnetic field gradient (taken normal to the surface) $\braket{\nabla B_{ecr}}$, averaged on the closed cold ECR surface~\cite{Li20}.
At the energies of interest, the photon range in condensed matter reaches several cm and a significant part of the flux is stopped within the superconducting magnet cold mass surrounding the plasma chamber. Photon power deposited in the cold mass as high as 1 W per kW of injected microwave power has been reported in the literature~\cite{Lyneis06}. A recent bremsstrahlung spectrum measurement, performed with the Jyväskylä A-ECR ECRIS~\cite{Bhaskar21}, having the rare possibility to observe the plasma both radially and axially, confirmed previous results reporting that a radial x-ray spectral temperature higher than the axial one~\cite{Lamoureux_eibrem,Noland2010}. These results are understood as a consequence of the presence of high magnetic field in ECRIS, completed with the existence of a hot relativistic electron population characterized by an anisotropic bremsstrahlung angular emission~\cite{Thuillier_bremsstrahlung}.
\\
LPSC is currently involved in the design of a new 28~GHz superconducting ion source named ASTERICS~\cite{Simon23,Thuillier23}. 
The expectation of significant parasitic x-ray emissions from this new device drives the need for a numerical study, for the first time, of the bremsstrahlung volume emission from an ECRIS plasma. An existing Monte Carlo (MC) code, used to track electrons in ECRIS~\cite{Thuillier_VenusInvestigation}, has been adapted to the ASTERICS design and completed with new functionalities to investigate volume bremsstrahlung emission from the ECRIS plasma. This study aims to better understand the x-ray generation in ECRIS as a function of the magnetic field configuration and find correlations with the plasma electron population properties, measured by the MC code with high statistics. 
After a brief presentation of the ASTERICS ECRIS geometry and magnetic structure (Section~\ref{sec:asterics}), the Monte Carlo model used in this work is presented (Section~\ref{sec:MonteCarlo}) and the simulation objectives and method are explained. Afterward, the simulation results are detailed and discussed in Section~\ref{sec:simulation}. Finally, the conclusions of the work are given (Section~\ref{sec:conclusion}).


\section{ASTERICS ion source}
\label{sec:asterics}

The ASTERICS ion source is a new superconducting ECRIS under design in France for the new GANIL injector (NEWGAIN) project~\cite{Ackermann21}, aiming at developing a new mass over charge $\leq$ 7 injector to the existing SPIRAL2 linear accelerator at the Grand Accelerateur National d'Ions Lourds (GANIL) facility.
The magnetic field will be generated by a set of 3 axial superconducting magnets and a hexapole coil. Details on the superconducting magnet and its cryostat are available in~\cite{Simon23}, while information on the ion source design and discussion on its expected performances are reported in~\cite{Thuillier23}. The main ASTERICS design parameters are summarized in Table~\ref{tab:astericsparameters}.

\begin{table}[b]
\caption{\label{tab:astericsparameters}
ASTERICS ECRIS main parameters.
}
\begin{ruledtabular}
\begin{tabular}{lcr}
Parameter & Value & Unit\\
\colrule
ECR operating frequency & 28 & GHz\\
$B_{ecr}$ & 1 & T\\
Maximum $B_{inj}$ & 3.7 & T\\
Minimum $B_{min}$ & 0.1 & T\\
Maximum $B_{ext}$ & 2.5 & T\\
Maximum $B_{hex}$ & 2.4 & T\\
Minimum $B_W$ & 2.0 & T\\
Peak to peak axial mirror length & 600 & mm\\
Plasma chamber inner radius & 91 & mm\\
\end{tabular}
\end{ruledtabular}
\end{table}

The ion source has a plasma chamber of 600 mm length and a 91 mm radius. The plasma chamber volume is larger than the other existing superconducting ECRIS, in order to enhance the achievable ion beam intensities. The ion source is designed to operate at 28~GHz, with a closed ECR resonance surface at 1 T and a last closed iso-B surface at the wall $B_W\approx 2$ T. The minimum-B magnetic field structure follows the standard magnetic scaling laws~\cite{Hitz02}, with a maximum axial magnetic mirror field profile of 3.7 T - 0.1 T - 2.5 T and a radial field intensity at the plasma chamber wall up to 2.4 T. The maximum magnetic profile of the ASTERICS source is displayed in Fig.~\ref{fig:B_asterics}.


\section{Monte Carlo code}
\label{sec:MonteCarlo}
The large ASTERICS ion source volume (15.6 dm$^3$) and the long timescale required to study the electron dynamics ($\approx$ 1 ms), combined with the presence of a hot and non-collisional electron population in ECRIS (estimated at $kT_e\approx$ 5-10 keV), drives us to use a Monte Carlo (MC) code to investigate the electron dynamics in the plasma, rather than a particle-in-cell code. An existing MC particle-tracking code~\cite{Thuillier_VenusInvestigation}, able to simulate the electron behavior inside ECRIS, has been adapted to the ASTERICS ion source geometry.
The field maps have been generated with either Poisson Superfish~\cite{POISSONSUPERFISH} or the OPERA simulation software~\cite{Opera_software}. The fringe field of the hexapolar magnetic field has been modeled with an analytical formulation~\cite{thuillier20}.
The electron pusher exploits the relativistic Boris scheme \cite{Boris_pusher}, known for its high stability. The radio-frequency (RF) electromagnetic field is modelled by a circularly polarized transverse planar wave, propagating along the plasma chamber axis $z$. The electric field intensity of the wave is matched to a given input power, whose value is discussed later in the text. The effect of the plasma potential is neglected.
The ions in the plasma are modelled as a static and homogeneous argon ion population background, present in the whole cavity. The plasma is assumed to be under-dense, reaching 15\% of the cutoff density (i.e. $n_e\approx1.6\cdot10^{18}\;m^{-3}$) for the injected 28~GHz RF. The ion charge state distribution (CSD) has been modelled from experimental measurements taken with the PHOENIX-V2 ECRIS, with an average ion CSD value of $Z_{av}=7.5$ and a peak particle intensity on Ar$^{10+}$.
\\
The electron collisions considered in the MC are Coulomb scattering, electron impact and bremsstrahlung emission. The electron radiative recombination is neglected, as it requires low energy electrons which are not studied in this model, focusing on hot electrons generated by the ECR mechanism. The electron radio-frequency scattering (see~\cite{Girard98}), which is the main candidate for the high energy electron deconfinement, is not modelled either in order to maximize the electron lifetime and the bremsstrahlung photon emission rate. The 40 kV high voltage extraction system of the ion source is modelled by a purely axial electric field of 10 kV/cm applied along a 4 cm distance, located right after the extraction plane. Electrons escaping through the extraction hole of 10 mm diameter would enter this repelling electric field region. A biased disk electrode (with a diameter of 20 mm), enhancing the cold electron confinement and located on the injection side, is set to the negative potential of -100 V. The associated electric field is modelled by a simple uniform axial field expanding on 1 mm, corresponding to the reduced distance to consider the electric field plasma screening.


\subsection{Coulomb collisions}

The Coulomb collision is modelled as a frequent and constant small scattering angle $\delta\theta$ changing randomly the electron velocity direction~\cite{Freidberg_book}. $\delta\theta$ is a function of the employed time-step and of both the electron-electron and electron-ion Coulomb collision frequencies (labelled respectively $\overline{\nu _{ee}}$ and $\overline{\nu _{ei}}$):

\begin{eqnarray}
\overline{\nu _{ei}}=\frac{\sqrt{2}}{12\pi^{\frac{3}{2}}}
\frac{e^4\ln\Lambda_e}{\epsilon_0^2 m_e^{1/2}(kT_e)^{3/2}}\sum_{i=Ar^{1+}}^{Ar^{16+}}n_iZ_i^2\;,
\\
\overline{\nu _{ee}} \approx \frac{\overline{\nu _{ei}}n_e}{Z_{av}^2 n_i}\;,
\label{eq:coll_freq}
\end{eqnarray}

where $n_i$ and $n_e$ are respectively the ion and electron number density, $e$ is the elementary charge, $\epsilon_0$ is the vacuum permittivity, $m_e$ is the electron mass and $\ln\Lambda_e$ is the Coulomb logarithm (set to $\approx$ 10 in this work). $T_e$ is the average electron temperature, considered equal to $kT_e\approx$ 5 keV in this work.


\subsection{Electron impact ionization}

Electron impact ionization of energetic electrons on atoms and ions is the main mechanism responsible for increasing the ions' charge state in ECRIS~\cite{Geller_book}. The collision induces a change in the incoming electron momentum that can possibly deconfine it from the ECRIS magnetic bottle. The cross-section for this physical phenomenon has been evaluated for all argon charge states using the Lotz semi-empirical formula~\cite{Lotz_crsect,Lotz_empformula}, which can approximate the experimental values within a 10\% error for the ions of interest.
The rate of electron impact ionization inside the plasma is handled with the Null-collision method~\cite{Hafi_3ViewNullColl,Koura_directNullColl,Koura_improvedNullColl}.
When an electron-ion collision occurs, the electron is assumed to lose an energy in the order of a few times the ion's ionization potential, and it is scattered with a random angle complying with the global collision momentum conservation. The kinematics model assumes an inelastic collision between the impinging and generated electrons. 


\subsection{Bremsstrahlung emission}

A new module has been added to the MC code, allowing to generate bremsstrahlung photons created by the interaction of the tracked electron with the ion and electron background population present in the plasma volume. The semi-empirical differential cross-section (DCS) of bremsstrahlung as a function of the photon energy $W$ is

\begin{equation}
 \label{eq:brDCS}
 \frac{d\sigma_{brem}}{dW}(T,Z)=\frac{Z^2}{\beta^2 W} \chi (Z,T,\kappa) \; ,
\end{equation}
where $Z$ is the atomic number of the target, $\beta$ is the electron velocity normalized to the speed of light, and $\kappa$ is the ratio between the photon energy $W$ and the electron kinetic energy $T$. $\chi$ is the scaled energy-loss differential cross-section, whose values are tabulated (see Appendix of \cite{Thuillier_bremsstrahlung} for context and \cite{Penelope_book,Seltzer_bremES} for details).

\begin{figure}[b]
\includegraphics[width=0.48\textwidth]{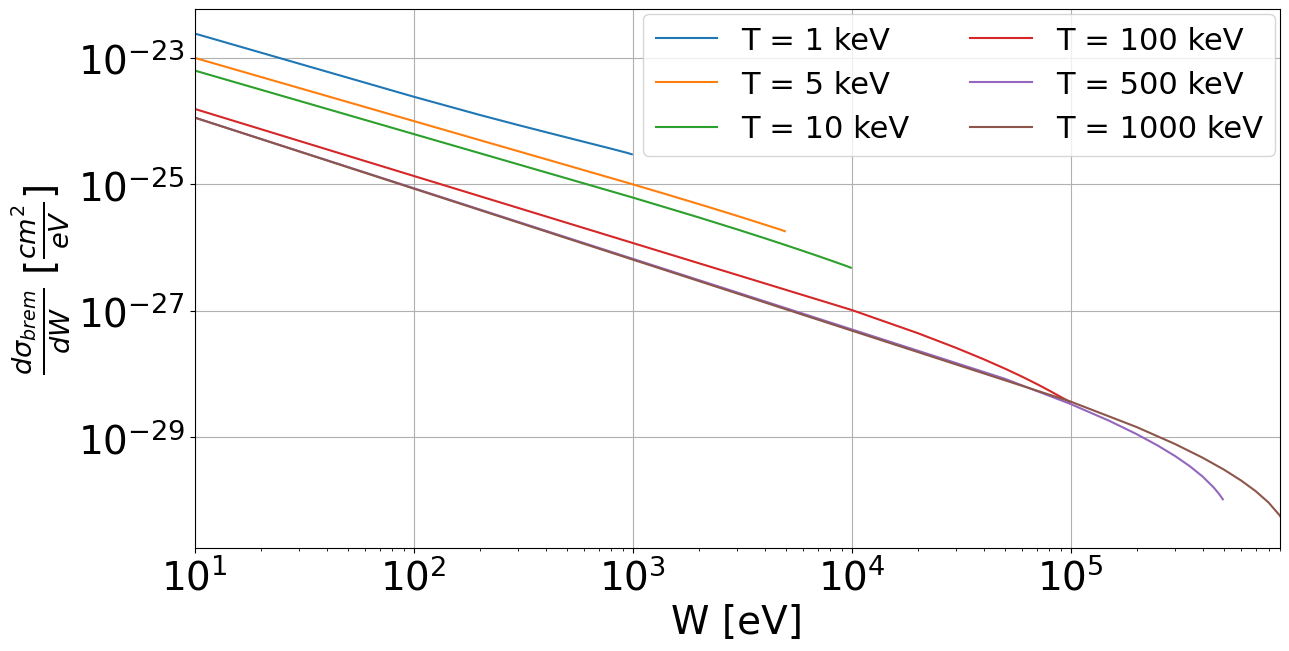}
\caption{\label{fig:DCS} Bremsstrahlung differential cross-section of electrons impinging on an argon ion, as a function of the emitted photon energy for different values of electron kinetic energy responsible for bremsstrahlung emission.}
\end{figure}

The double bremsstrahlung DCS is
\begin{equation}
 \label{eq:brDDCS}
 \frac{d^2\sigma_{brem}}{dWd\cos{\theta}}=\frac{d\sigma_{brem}}{dW}p_{brem}(Z,T,\kappa,\cos{\theta}) \; ,
\end{equation}

where $p_{brem}$ is the shape function, whose tabulated values can be found in~\cite{Kissel_shapefunc}. Examples of DCS for different electron energies impinging on an argon plasma are represented in Fig.~\ref{fig:DCS}.

The implementation of the bremsstrahlung emission in the MC is similar to the Penelope code~\cite{Penelope_book}, which samples a direction and energy value for the emitted photons starting from a probability distribution function built upon tabulated values of scaled energy loss DCS and shape function.
The tabulated function values used in this work are taken from the Poškus code~\cite{Poskus_shapefunc}, able to calculate both $\chi$ and $p_{brem}$ on a much thinner grid with respect to other sources from the literature~\cite{Seltzer_bremES,Kissel_shapefunc}.
The Null-collision method~\cite{Hafi_3ViewNullColl,Koura_directNullColl,Koura_improvedNullColl} has been employed to include the dependencies for photon emission on the bremsstrahlung cross-section and electron velocity.
The total bremsstrahlung cross-section $\sigma_{tot}(T)$ and collision rate $n_i\cdot\sigma_{brem}\cdot v(T)$ are proposed in Table~\ref{tab:bremsstrahlung_data} for a few selected electron kinetic energies (T = 10 keV, 100 keV and 1000 keV). One can see that $\sigma_{tot}$ is dominant at lower electron kinetic energy (441 barn for T = 10 keV) and decreases from 75 barns down to tens of barn for 100 keV and 1 MeV electrons respectively. The expected bremsstrahlung collision rate is 0.5 Hz for 10 keV electrons and reduces to 0.16 Hz at 1 MeV. These very low collision rate frequencies give rise to modelling difficulty in the MC code. This point is extensively discussed in the next subsection. 

\begin{table}[t]
\caption{\label{tab:bremsstrahlung_data}%
Bremsstrahlung data for total cross section ($\sigma$) and collision rate ($n_i\cdot v\cdot\sigma$) at different kinetic energy (T) of the electron emitting the bremsstrahlung radiation. $v$ is the associated electron velocity modulus.
}
\begin{ruledtabular}
\begin{tabular}{ccc}
\textrm{T [keV]}&
\textrm{$\sigma_{tot}$ [barn]}&
\textrm{$n_i\cdot \sigma_{brem} \cdot v$ [Hz]}\\
\colrule
10 & 441.82 & 0.50\\
100 & 75.37 & 0.24\\
1000 & 28.83 & 0.16\\
\end{tabular}
\end{ruledtabular}
\end{table}


\subsection{Simulation objectives and method}

\begin{table}[b]
\caption{\label{tab:axial_profiles}%
Ion source magnetic field values (in Tesla) employed in the simulation for the 9 magnetic configurations studied, named $C_1$, $C_2$, ... up to $C_9$. The magnetic parameters in the table are explained in Section~\ref{sec:introduction} and represented in Fig.~\ref{fig:B_asterics}.}
\begin{ruledtabular}
\begin{tabular}{cccccccc}
\textrm{$B_{inj}$}&
\textrm{$B_{min}$}&
\textrm{$B_{ext}$}&
\textrm{$B_{hex}$}&
\textrm{$B_{Winj}$}&
\textrm{$B_{Wext}$}&
\textrm{$R$}&
\textrm{Label}\\
\colrule
3.7 & 0.3 & 2.0 & 2.4 & 2.04 & 2.09& 6.8&$C_1$\\
3.7 & 0.3 & 2.5 & 2.4 & 2.02 & 2.01& 6.7&$C_2$\\
3.7 & 0.3 & 2.2 & 2.4 & 2.03 & 2.06& 6.8&$C_3$\\
3.7 & 0.4 & 2.2 & 2.4 & 2.09 & 2.10& 5.2&$C_4$\\
3.7 & 0.5 & 2.2 & 2.4 & 2.14 & 2.15& 4.3&$C_5$\\
3.7 & 0.6 & 2.2 & 2.4 & 2.18 & 2.21& 3.6&$C_6$\\
3.7 & 0.7 & 2.2 & 2.4 & 2.24 & 2.26& 3.1&$C_7$\\
3.7 & 0.8 & 2.2 & 2.4 & 2.29 & 2.31& 2.8&$C_8$\\
3.7 & 0.9 & 2.2 & 2.4 & 2.36 & 2.37& 2.4&$C_9$\\
\end{tabular}
\end{ruledtabular}
\end{table}

The objectives of the present electron MC simulation study are to:
\begin{itemize}
\item investigate with a high electron statistics the effect of the ECRIS magnetic field configuration on the electron density, energy, and both the parallel and perpendicular velocity distributions;
\item investigate the effect of the ECRIS magnetic field on the bremsstrahlung emission of electrons impacting ions in the plasma volume.
\end{itemize}

Nine axial magnetic mirror profiles, summarized in Table~\ref{tab:axial_profiles}, have been studied with the MC simulations. Note that the hexapole magnetic field intensity has been kept constant and equal to $B_{hex}$ = 2.4 T for all the simulations. The axial profile labelled $C_3$ is considered as the base configuration, with $B_{min}$ = 0.3 T and $B_{ext}<B_{hex}$, allowing for an optimized ion beam extraction from the source and the injection of both 18 and 28~GHz microwaves (the double frequency heating has not been implemented in this work for the sake of simplicity). Such configuration is experimentally known to provide a stable plasma with a low bremsstrahlung spectral temperature~\cite{Lyneis06}. 
The comparison of the cases $C_1$, $C_3$ and $C_2$ enables investigating the effect of larger extraction peak magnetic fields, with $B_{ext}$ values of 2.0 T, 2.2 T and 2.5 T respectively, keeping the other magnetic parameters constant.
On the other hand. the cases $C_3$, $C_4$, $C_5$ ... $C_9$ allow studying the specific effect of increasing $B_{min}$ values with the other magnetic parameters kept constant. 
\\
The standard condition of simulation assumes a total injected RF power of 7 kW at 28~GHz. The electrons are uniformly generated inside the ECR zone surface with a uniformly distributed random velocity direction. The initial electron energy is cast assuming electrons are created from the ion CSD described earlier.
For a specific charge state $i$, the electron initial kinetic energy is randomly chosen from a Gaussian distribution, whose mean and standard deviation are the ionization potential $I_i$ and $0.1\times I_i$ respectively.
The simulated electrons are propagated until they get deconfined from the plasma chamber, or when the propagation time reaches 1 ms. This time limit represents a balance between a realistic electron lifetime and a reasonable computational cost, being approximately two weeks for an $1.25\times 10^6$ electron statistics, using 80 parallel processes. During the MC simulations, the initial and final states of each electron are saved. The local electron parallel and perpendicular velocity (with respect to the local magnetic field direction) distributions, the electron relative density and the electron energy distribution function (EEDF) are recorded inside a set of spatial planar grid with a 1 mm x 1 mm precision, located in specific places in the plasma chamber, using histograms incremented each time an electron passes through them.
\\
The MC bremsstrahlung simulation aims at investigating the x-ray spectrum with energies in the range $\approx$ 0.15 – 1 MeV. This requires to record a sufficient statistics of high energy photons. Considering the actual collision rates of bremsstrahlung (see Table~\ref{tab:bremsstrahlung_data}) and the maximum propagation time for the electrons (1 ms), one can see that reaching this goal with the actual physical bremsstrahlung parameters would require a huge electron statistics in the order of $\approx 10^{10}$, and a total prohibitive computing time of approximately 400 years with the 80 processors available. In order to reach the required statistic within a reasonable time of a few weeks, the following adaptations have been implemented in the MC:
\begin{itemize}
 \item artificially boost the bremsstrahlung cross-section by a factor $10^4$;
 \item increase the electron limit propagation time from 1 ms to 2 ms;
 \item boost the RF electric field by a factor of 2 (to reach 20 kV/m), in order to favor the high energy electrons production.
\end{itemize}

\begin{table}[b]
\caption{\label{tab:telescope}%
Cylindrical coordinates and geometry definition of the axial and radial telescope characteristics, in both the nominal and amped simulation setup. The definition of the parameters is explained in Fig.~\ref{fig:telescope}.
}
\begin{ruledtabular}
\begin{tabular}{ccc}
Parameters & Radial telescope & Axial telescope \\
\colrule
\textrm{$z_1$} [mm] & 3 & 303 \\
\textrm{$z_2$} [mm] & 63 & 803\\
\textrm{$r_1$} [mm] & 91 & 0 \\
\textrm{$r_2$} [mm] & 591 & 27.3\\
\textrm{$L_z$} [mm] & 60 & 500 \\
\textrm{$L_r$} [mm] & 500 & 27.3\\
\textrm{Surface} [mm$^2$] & 34206 & 2341\\
\textrm{$\Omega$} [sr] & 0.637 & 0.0035\\
\end{tabular}
\end{ruledtabular}
\end{table}

\begin{figure}[t]
\includegraphics[width=0.48\textwidth, valign=t]{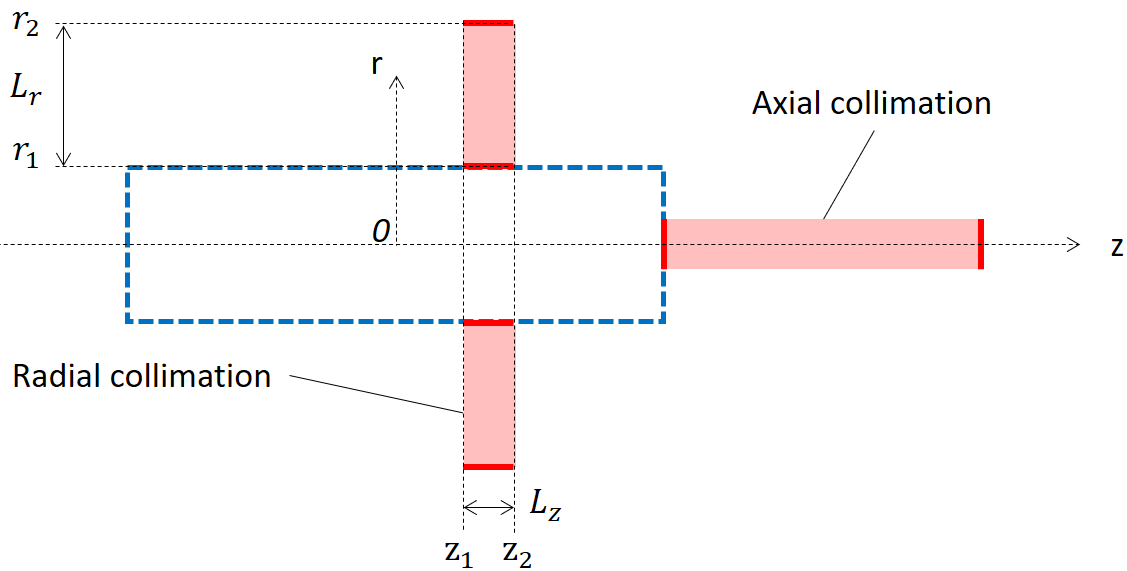}
\caption{Cylindrical parameters defining the geometry of the axial and radial collimators considered to investigate the bremsstrahlung emission in the simulation. The plasma chamber limits are displayed with a dashed blue line (cylinder with -297 mm $\leq$ z $\leq$ 303 mm and r $\leq$ 91 mm)}.
\label{fig:telescope}
\end{figure}

The electric field intensity of 20 kV/m corresponds to a 14 kW injected RF power, which is a physical value for a 15 liter plasma chamber volume (usual existing ECR ion sources operate with a RF power density of the order of 1 kW/l). Moreover, applying the same RF settings to all the magnetic configurations should allow a fair comparison of the results relatively one to each other. 
\\
The methodology to generate the x-ray spectrum is achieved as follows. First, for each magnetic field configuration, a set of $10^5$ electrons is randomly generated and propagated up to 2 ms, without enabling bremsstrahlung emission. Any electron reaching 150 keV is stopped and its final position, velocity and final RF phase are saved. The output of this first step is a file containing a collection of randomly generated electrons which reached a kinetic energy of 150 keV. This electron collection is used as a future seed to boost the high energy electron tail x-ray production.
The second step consists in generating $10^5$ electrons randomly selected from those output file, for each magnetic configuration. Bremsstrahlung emission is enabled in these simulations, and x-rays with energy above 150 keV are saved. Since the electrons undergo Coulomb collision, they quickly lose the memory of their origin and each generated electron provides a unique photon spectrum due to its random energy history. The data recorded during the bremsstrahlung runs are:

\begin{enumerate}
    \item The location of photon crossing the cylindrical plasma chamber wall, stored in a 40 bin histogram. There are 20 uniformly distributed bins covering the plasma chamber radial wall from z = -0.3 m to z = 0.3 m. Each bin counts photons passing through a cylinder of axial length $dz$ = 30 mm and radius 91 mm. Moreover, 10 concentric disks cover both the injection and extraction walls with a constant radial thickness $dr$ = 9.1 mm.
    \item The energy spectrum of photons in the range 150 to 2000 keV, with a 10 keV resolution. The photon energy spectrum is registered in specific solid angles, hereafter referred to as telescopes, displayed in Fig.~\ref{fig:telescope} and whose geometrical parameters are detailed in Table~\ref{tab:telescope}.
    \item The full 4$\pi$ photon energy spectrum, recorded in a dedicated histogram. 
\end{enumerate}

The axial telescope is composed of two axi-symmetric disks separated by a distance of 500 mm, with inner and outer radius of 0 and 27.3 mm respectively. The first surface layer is set at the plasma extraction wall ($z=0.3$ m), while the second telescope layer is located at $z=0.8$ m.
The radial telescope is composed of two concentric cylindrical surfaces located above the axial position of $B_{min}$ ($ 0\leq z \leq 60$ mm), with radii of 91 mm and 591 mm respectively. Each photon passing through these two layers is counted in a dedicated histogram.
\\
Each $10^5$ electron statistics is split into 100 jobs of $10^3$ electrons each, sent in parallel on a Linux server. The individual job histograms are later merged together to ease the data analysis. Since the axial photon statistics is small with respect to the others, five runs of each configuration have been necessary to reach a statistics sufficient to do the analysis. The typical duration of a bremsstrahlung job is approximately between 1 and 2 days on each processor.

\section{Simulation results for electron}
\label{sec:simulation}


\subsection{Electron density distribution}
\label{sec:density}

\begin{figure}[t]
\centering
\includegraphics[width=0.45\textwidth]{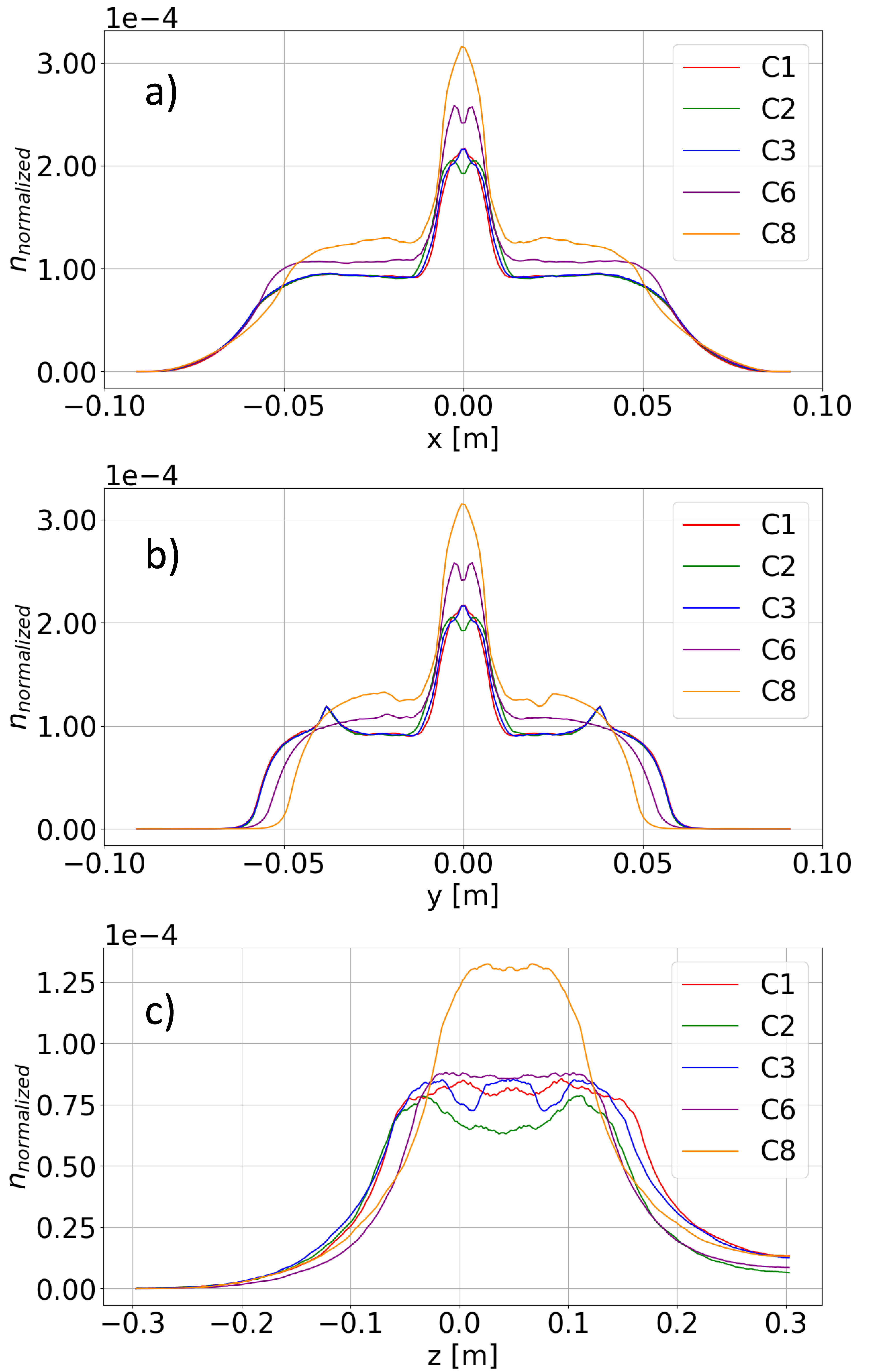}
\caption{Electron density distribution profiles for the magnetic configurations $C_1$, $C_2$, $C_3$, $C_6$ and $C_8$ \textbf{a)} in the XY plane at B$_{min}$ along x-axis with y=0; \textbf{b)} in the XY plane at B$_{min}$ along y-axis with x=0; \textbf{c)} in the XZ plane along z-axis at y = 0. \textbf{N.B.} the normalization factor is identical for figures \textbf{a)} and \textbf{b)}  and different for figure \textbf{c)}.}
\label{fig:density}
\end{figure}

The high statistics of the performed MC simulation has allowed to generate high resolution normalized electron density maps of the plasma inside the plasma chamber. Representative density profiles for the cases from $C_1$, $C_2$, $C_3$, $C_6$ and $C_8$ are reported in Fig.~\ref{fig:density}, taken along the 3 main axis of the plasma chamber (X, Y and Z for subplots (a), (b) and (c) respectively) and passing by the axis point where $z=z_{min}$. Two dimension (2D) density plots for cases $C_3$ and $C_8$ are presented in the Appendix (Fig.~\ref{fig:n_densityplot_long}) for completeness.
The presence of a dense electron population, matching the ECR zone volume and mainly concentrated in a compact cylinder centered around the ion source axis, is systematically observed for all the magnetic cases studied.
The obtained results are comparable to the one reported in previous works~\cite{Heinen_1999,Mascali_3Dsymm,Mironov_2021}, characterized by self-consistent calculations including ion dynamics, plasma potential and/or RF mode coupling dynamic calculations. This demonstrates that the hot electron population of ECRIS is marginally affected by the plasma potential and the RF field cavity modes induced by the plasma, which  validates the hypothesis considered in this study, consisting in  using a simple MC code to investigate the hot electrons behaviour in ECRIS plasma.
\\
It is worth noting that when $B_{ext}$ is varied in Fig.~\ref{fig:density}, the electron density distribution remains practically unchanged (cases $C_1$, $C_2$, $C_3$), while the electron density profiles shrinks in size and increase in intensity both axially and radially for growing values of $B_{min}$ (see plots for cases $C_3$, $C_6$ and $C_8$). These changes closely follow the subsequent reduction of the ECR surface dimension, resulting in higher local electron densities (see also the ECR zone plotted in black dashed lines in Fig.~\ref{fig:n_densityplot_long} in the Appendix). When the ECR zone shrinks in size, magnetically trapped electrons close to the source axis bounce on shorter distances and are therefore more concentrated in space. Indeed, the peak electron density intensity in Fig.~\ref{fig:density}(a) and (b) increases by approximately 50\% when $B_{min}$ changes from 0.3 T to 0.8 T (cases $C_3$ and $C_8$). This increase is also visible along the z direction in subplot~\ref{fig:density}(c).
\\
In the configuration $C_8$ ($B_{min}$ = 0.8 T), the central density peak full width half maximum (FWHM) is $\approx$ 17 mm for the ASTERICS source, while it is $\approx$ 7 mm for the same $B_{min}$ in DECRIS-PM~\cite{Mironov_2021}. These distances, normalized to the respective plasma chamber radius, become very similar ($\approx$9.3\% and $\approx$10\% respectively). In ECRIS, the locations characterized by denser electron population are commonly associated with the presence of multicharged ions. This proportional relation between the size of the central dense electron cloud and the radius of the plasma chamber, considering the usual ECRIS magnetic scaling laws for the magnetic field~\cite{Hitz02}, is a simulation confirmation of the project choice to design the new ASTERICS ion source with a large plasma chamber, in order to increase the ion population in the source and the potential ion beam intensities~\cite{Thuillier23}.
\\
Another result visible in Fig.~\ref{fig:density}(c) is a significant drop (-33\%) in the electron longitudinal density nearby the plane of ion beam extraction (z $\approx$ 0.3 m), when $B_{ext}$ is varied from 2.2 T ($C_3$) to 2.5 T ($C_2$). This decrease can be related to the specific magnetic configuration of $C_2$, characterized by $B_{ext} > B_{hex}$ and $B_{ext} \gg B_W$. The high $B_{ext}$ intensity repels a larger fraction of electron population through magnetic mirroring, which will get preferentially deconfined on the other side of the magnetic field lines connected to the extraction (namely the radial wall where $B_{W}\approx 2.0$ for $C_2$).


\subsection{Electron Confinement Time and Losses}
\label{sec:loss}

\begin{figure}[t]
\includegraphics[width=0.48\textwidth]{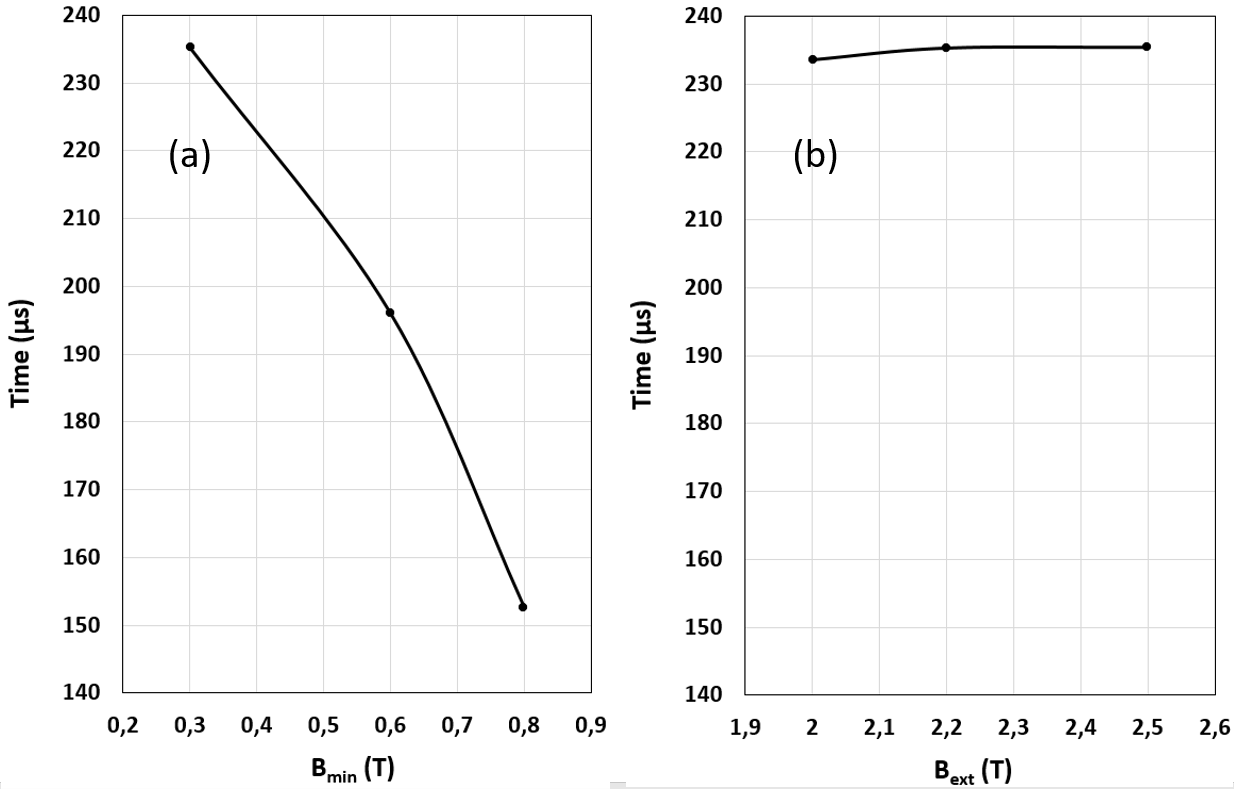}
\caption{Averaged electron confinement time for particles surviving more than 10 $\mu$s in the cavity as a function of $B_{min}$ (subplot a) and $B_{ext}$ (subplot b).}
\label{fig:conftime}
\end{figure}

Figure~\ref{fig:conftime} presents the simulated averaged electron confinement time (evaluated considering the electron population deconfined at the chamber walls) as a function of $B_{min}$ (Fig.~\ref{fig:conftime}.a) and $B_{ext}$ (Fig.~\ref{fig:conftime}.b). The electrons with lifetimes below 10 $\mu$s, corresponding to low-energy particles which are quickly deconfined, have been excluded from the average. When $B_{min}$ is increased from 0.3 T to 0.8 T, the averaged confinement time drops from 233~$\mu$s to 152 $\mu$s.
This drop is well explained as a general decrease of the axial and radial magnetic mirror confinement: the characteristic time for electrons to be Coulomb scattered to the wall is $\delta t \approx \Delta \theta^2$, where $\Delta \theta=\pi/2-asin(1/\sqrt{R})$ and $R$ is the magnetic mirror ratio.
From Table~\ref{tab:axial_profiles}, it can be observed that $R$ decreases from 6.8 for $B_{min}$ = 0.3 T to 2.8 for $B_{min}$ = 0.8 T: the confinement time drop calculated with the above formula is $-38\%$, while the simulation gives a reduction of $-35\%$. 
The confinement time constancy observed  when $B_{ext}$ varies (in the order of 230 $\mu$s) has the same explanation, since $R$ is approximately constant when $B_{ext}$ is changed from 2 T to 2.5 T (see Table~\ref{tab:axial_profiles}).

\begin{figure}[b]
\includegraphics[width=0.45\textwidth]{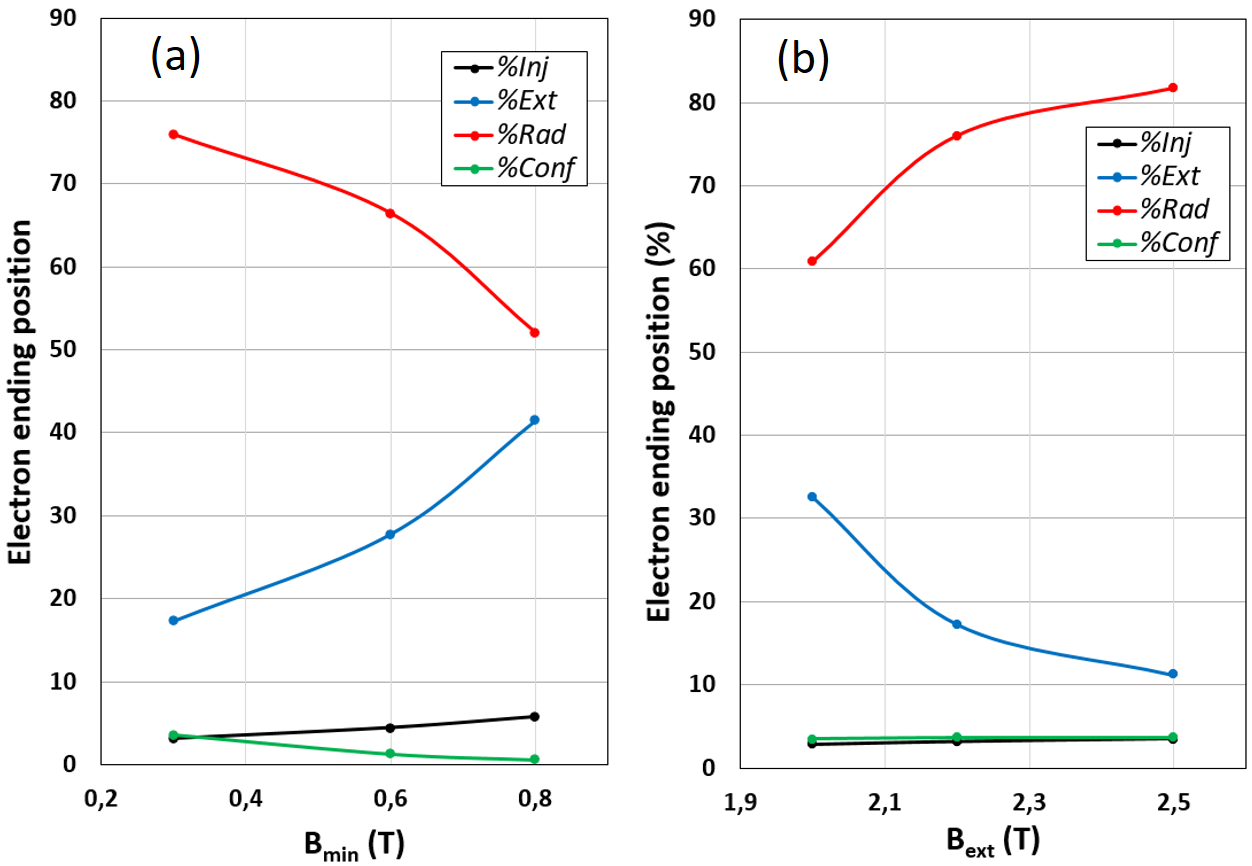}
\caption{Normalized relative evolution of the four electrons' ending positions ($\%_{Inj}$, $\%_{Ext}$, $\%_{Rad}$ and $\%_{Conf}$) in the simulation as a function of $B_{min}$ and $B_{ext}$, on subplot (a) and (b) respectively. See text for details.}
\label{fig:loss}
\end{figure}

The electrons' final position at the end of the propagation time limit (1 ms) are classified into 4 normalized groups, whose name and definition are as follows.

\begin{itemize}
    \item $\%_{Inj}$: electrons hitting the injection plate (trans\-versal plane at z = -0.3 m);
    \item $\%_{Ext}$: electrons hitting the extraction electrode (transversal plane at z = 0.3 m);
    \item $\%_{Rad}$: electron hitting the radial wall (cylindrical shell with inner radius r = 91 mm);
    \item $\%_{Conf}$: electrons still confined in the plasma volume.
\end{itemize}

The relative evolution of these 4 populations is plotted in Fig.~\ref{fig:loss}(a), as a function of $B_{min}$ (cases $C_3$, $C_6$ and $C_8$) and in Fig.~\ref{fig:loss}(b) as a function of $B_{ext}$ (cases $C_1$, $C_2$ and $C_3$). The magnetic field changes in the ECRIS have an important effect on the four aforementioned populations.
The opposite interplay  of $\%_{Ext}$ and $\%_{Rad}$ with variations of $B_{min}$ and $B_{ext}$ is a consequence of the relative evolution of the local magnetic mirrors $R_{ext}=B_{ext}/B_{min}$ and $R_{rad}=B_W/B_{min}$. Indeed, the Coulomb scattering model implemented in the MC code leads to a preferential electron deconfinement along the local field line end corresponding to the weakest magnetic field.
The general low loss count at injection can be explained by the much higher magnetic field intensity on this surface with respect to the other ones ($B_{inj}\gg B_W$ and $B_{inj}\gg B_{ext}$), efficiently repelling the incoming electrons through magnetic mirror. The increase of $\%_{inj}$ with $B_{min}$ is a consequence of the reduction of the injection mirror ratio $R_{inj}=B_{inj}/B_{min}$. The concomitant reduction of $\%_{Conf}$ can be associated to the already discussed confinement time reduction.
Finally, it is worth noting that variations of $B_{min}$ and $B_{ext}$ in Fig.~\ref{fig:loss} are such that the sum $\%_{Ext}+\%_{Rad}\sim const$, and are accounting for the majority of the losses.


\subsection{Electron Energy distribution}
\label{sec-simulation-Edistr}

\begin{figure}[b]
\centering
\includegraphics[width=0.48\textwidth]{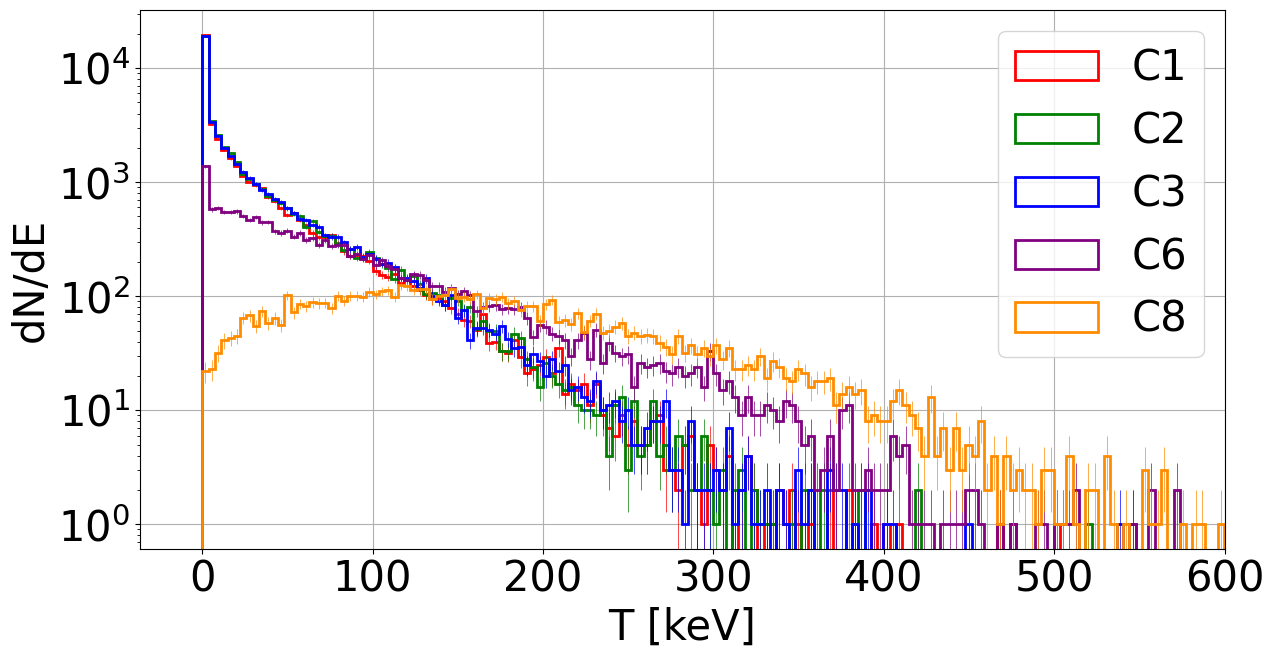}
\caption{Electron energy distribution function for the magnetic configurations  $C_1$, $C_2$, $C_3$, $C_6$ and $C_8$ of electrons still confined in the chamber volume at the end of the simulation.}
\label{fig:EEDF}
\end{figure}
 
\begin{figure}[b]
\centering
\includegraphics[width=0.48\textwidth]{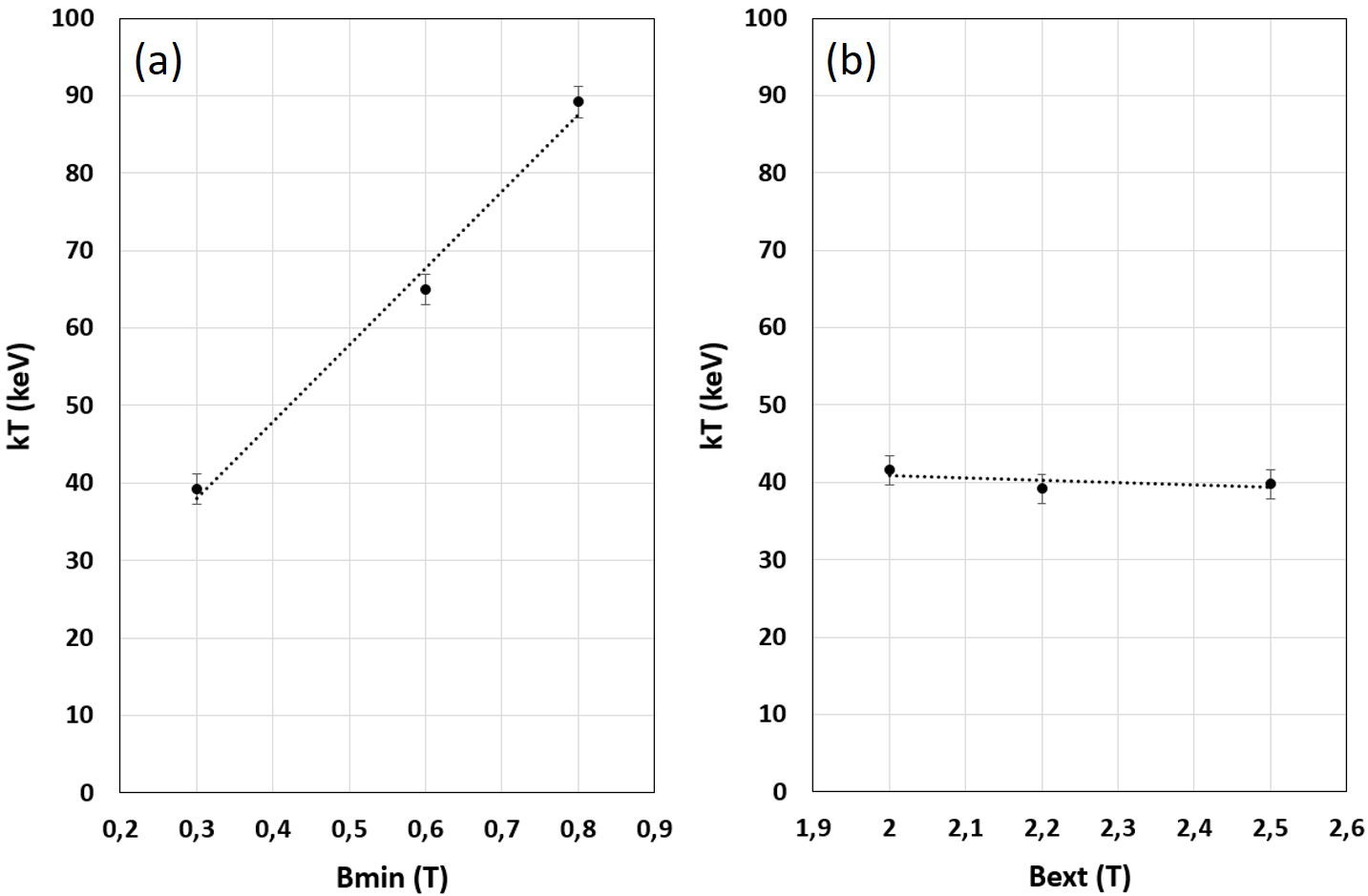}
\caption{Evolution of the high energy EEDF tail temperature for the electron population still confined in the source after 1 ms, as a function of $B_{min}$ (a) and $B_{ext}$ (b).}
\label{fig:kTe}
\end{figure}

Fig.~\ref{fig:EEDF} presents the EEDF of electron still confined in the volume of the plasma chamber at the end of the simulation (1 ms), for the magnetic configurations $C_1$, $C_2$, $C_3$, $C_6$ and $C_8$. The high energy part of these five EEDF is considered to represent the electron population providing the most significant contribution to the bremsstrahlung plasma emission. The hot electron energy tail of the EEDF, plotted in Fig.~\ref{fig:EEDF}, has been fitted with a single Maxwell-Boltzmann distribution function for electron kinetic energy $\geq$ 200 keV. The fit results are presented in Table~\ref{tab:T_hot_tail} and plotted against $B_{min}$ and $B_{ext}$ in Fig.~\ref{fig:kTe}(a) and Fig.~\ref{fig:kTe}(b) respectively.
Results for the confined electron EEDF are as follows:

\begin{table}[t]
\caption{\label{tab:T_hot_tail}
Electron temperature measured with the high energy tails of the EEDF presented in Fig.~\ref{fig:EEDF}, fitted with a Maxwell-Boltzmann distribution for the axial magnetic profiles from $C_1$, $C_2$, $C_3$, $C_6$ and $C_8$. The fit uncertainty is indicated in the last table line.}
\begin{ruledtabular}
\begin{tabular}{cccccc}
Case & $C_1$ & $C_2$ & $C_3$ & $C_6$ & $C_8$\\
\colrule
$kT_{e} (keV)$ & $41.6$ & $39.8$ & $39.2$& $65.0$ &$89.2$ \\
$\sigma_{kT_{e}} (keV)$ & $\pm 1.9$ & $\pm 1.4$ & $\pm 1.5$& $\pm 2.0$ &$\pm 1.6$ \\
\end{tabular}
\end{ruledtabular}
\end{table}

\begin{itemize}
 \item variations of $B_{ext}$ (case $C_1$, $C_2$ and $C_3$) do not change the hot electron tail temperature ($T\approx 40\; keV\approx const$, see Fig.~\ref{fig:kTe}(b));
 \item the $B_{min}$ increase results in a quasi-linear growth of the hot electron tail temperature (respectively 39 keV, 65 keV and 89 keV for $B_{min}$ = 0.3 T, 0.6 T and 0.8 T, see Fig.~\ref{fig:kTe}(a));
 \item the EEDF for $B_{min}=0.8$ T is featuring a positive slope up to $\approx$ 150 keV, followed by a negative one (see Fig.~\ref{fig:EEDF}-$C_8$). 
\end{itemize}

It has been experimentally demonstrated that the onset of kinetic instabilities in ECRIS is observed for values of $B_{min}/B_{ecr} \geq 0.8$ T~\cite{Tarvainen14,Li20}. A condition to trigger kinetic instabilities in plasma is a change of slope in the EEDF. Hence, the EEDF obtained for the confined electrons at $B_{min}=0.8$ T is a possible simulation confirmation of the formation of an unstable ECR plasma, in accordance with former experimental works published.

\subsection{Electron Velocity Spatial Distribution}
\label{sec-simulation-vdistr}

\begin{figure*}[!t]
\centering
\includegraphics[width=0.95\textwidth, valign=t]{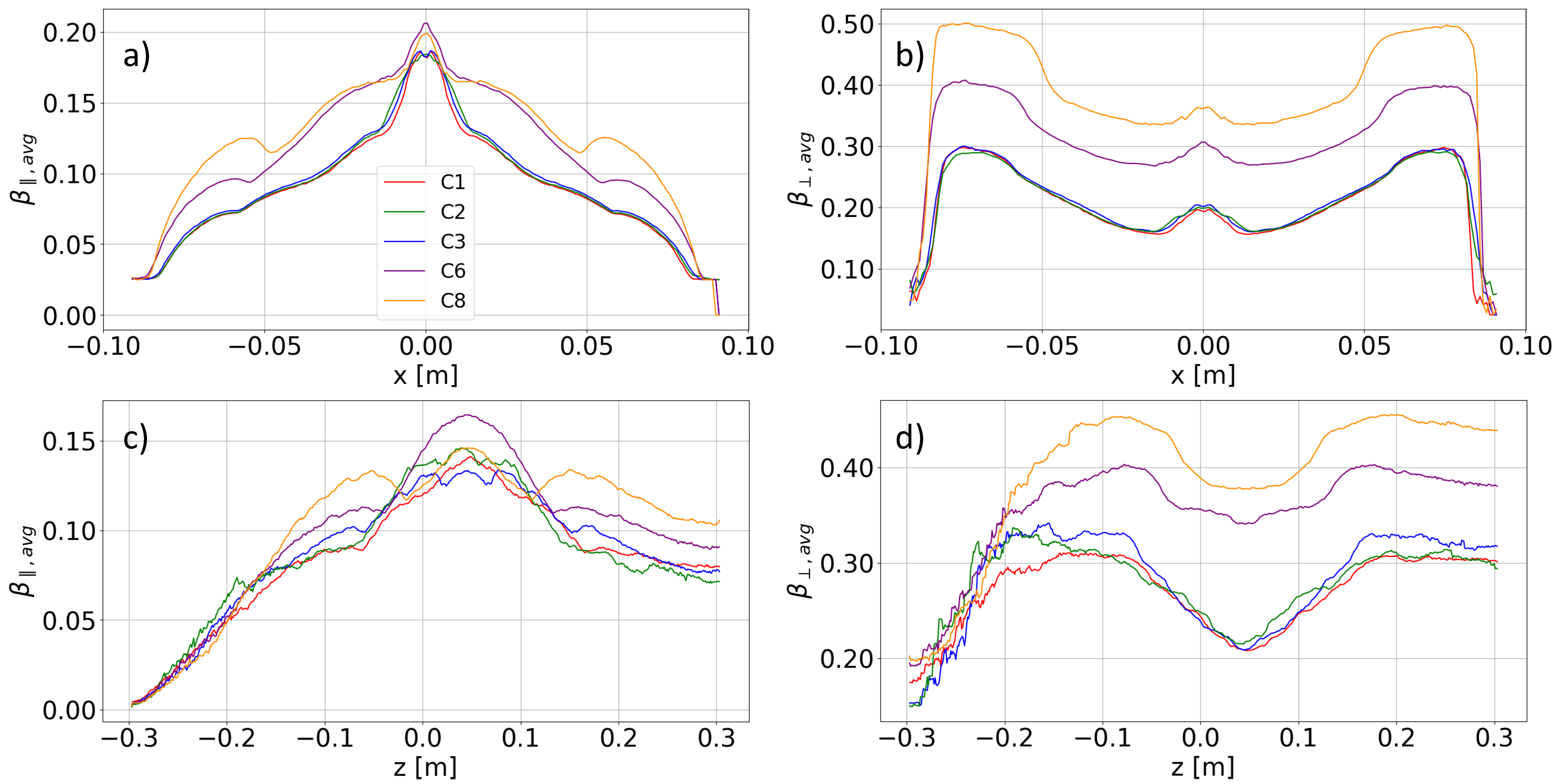}
\caption{Average parallel (left column, subplots (a) and (c)) and perpendicular (right column, subplots (b) and (d)) velocity distribution profiles for the magnetic configurations for $C_1$, $C_2$, $C_3$, $C_6$ and $C_8$, taken along the x-axis in the XY plane where $z=z_{min}$ (first row, subplots (a) and (b)) and along the z-axis at x=0 and y=0, (second row, subplots (c) and (d)).}
\label{fig:4_v}
\end{figure*}

The high statistics of the MC allows studying the spatial distribution of the averaged (absolute) velocity parallel ($\beta_{||}=v_{||}/c$) and perpendicular ($\beta_{\perp}=v_{\perp}/c$) to the local magnetic field inside the plasma chamber. 
 Fig.~\ref{fig:4_v} presents the evolution of $\beta_{||}$ (subplots (a) and (c)) and $\beta_{\perp}$ (subplots (b) and (d)) along two directions of interest for the magnetic cases $C_1$, $C_2$, $C_3$, $C_6$ and $C_8$. The plotting direction for the first row (subplots (a) and (b)) is the x-axis passing by $z=z_{min}$ in the center of the plasma, while the direction for the second row (subplots (c) and (d)) is taken along the z-axis.
It can be observed that when $B_{ext}$ is increased from 2.0 T to 2.5 T (cases $C_1$, $C_2$ and $C_3$), all the profiles remain almost unaltered. On the contrary, variations of $B_{min}$ (cases $C_3$, $C_6$ and $C_8$) have an important influence on the velocity profiles, as discussed below. On subplots (c) and (d), the adiabatic evolution of $\beta_{||}$ and $\beta_{\perp}$ inside the ECR zone is clearly visible, with $\beta_{||}$ and $\beta_{\perp}$ reaching respectively a maximum and a minimum at $B(z_{min})=B_{min}$.
\\
In Fig.~\ref{fig:4_v}(c), one can notice that the electrons reach the extraction plane ($z_{ext}$ = 0.3 m) with a high mean parallel velocity ($\approx 0.07-0.1$), while this behavior is not observed on the injection plane ($\beta_{||}\rightarrow 0$ when $z\rightarrow z_{inj}$= -0.3 m).
This trend is expected from magnetized electrons subjected to Coulomb scattering inside a magnetic mirror.
The large value of $\beta_{||}$  at $z=z_{ext}$ is a consequence of the electrostatic confinement imposed by the extraction electric field $E$.
Indeed, since $B_{inj} \gg B_{ext}$, almost all the axial electrons are naturally scattered toward the extraction plasma electrode. However, the extraction electric field repel them back into the plasma as long as their parallel kinetic energy is lower or equal to 40 keV, corresponding to $\beta_{||}=0.37$. This ability to repel electrons with such a large parallel velocity boosts the original extraction mirror ratio $R_{ext}=B_{ext}/B_{min}$ to much higher values. Simple calculations demonstrate that the actual  $R_{ext}(E) \gg R_{inj}$. Consequently, the majority of axially trapped electrons build up a parallel velocity much higher than $R_{ext}$ would allow until they are eventually deconfined.
The $\beta_{||}$ x-axis profile (Fig.~\ref{fig:4_v}(a)) is characterized by the presence of a central peak, concomitant with the density peak visible in Fig.~\ref{fig:density} (characterized by the radius $r_L$), surrounded by a progressive decrease to zero toward the plasma chamber radial wall. A change of slope is visible for $r \gtrsim$ 5 cm for all the magnetic configurations. This variation marks the presence of the ECR zone ($B=B_{ecr}$), where non-relativistic electrons rapidly increase their perpendicular velocity through resonant phenomena.
At $z=z_{min}$, the axial field intensity $B_{min}$ is quasi-constant in the transverse XY plane, and the ECR zone radius $r_{ecr}$ can be estimated as

\begin{equation}
r_{ecr} \approx r_W \left( \frac{B_{ecr}^2-B_{min}^2}{B_{hex}^2}\right) ^{1/4} \;.
\end{equation}

This formula yields $r_{ecr}$= 57 mm, 53 mm and 46 mm for $B_{min}$ = 0.3 T, 0.6 T and 0.8 T respectively. These values correspond indeed to the radius of the slope discontinuities in the respective curves ($C_1$, $C_2$, $C_3$), $C_6$ and $C_8$ in Fig.~\ref{fig:4_v}(a).
In the $\beta_{\perp}$ x-axis profiles (Fig.~\ref{fig:4_v}(b)), a small hump is visible for radius $r\leq r_L$. Above $r=r_{ecr}$, a steeper growth of $\beta_{\perp}$ is observed up to the plasma chamber wall $r_W$.
In the range $r_L \leq r \leq r_{ecr}$, $\beta_{\perp}$ smoothly increases following a power law, highlighting a correlation with the growth of the radial magnetic field intensity.
Indeed, in the plane $z=z_{min}$, the average (non-relativistic) electron magnetic moment $\mu_{av} \sim \frac{ m \beta_{\perp}^2c^2}{2B}$ display a quasi-constant behavior with $r$ for values within the limits $r_L \leq r \leq r_{ecr}$, leading to the approximate following equation for electrons

\begin{equation}
\label{eq:mub}\beta_{\perp}^2 \sim \sqrt{B_{min}^2+B_{hex}^2\frac{r^4}{r_W^4}}\;.
\end{equation}

\begin{figure*}
\centering
\includegraphics[width=0.95\textwidth, valign=t]{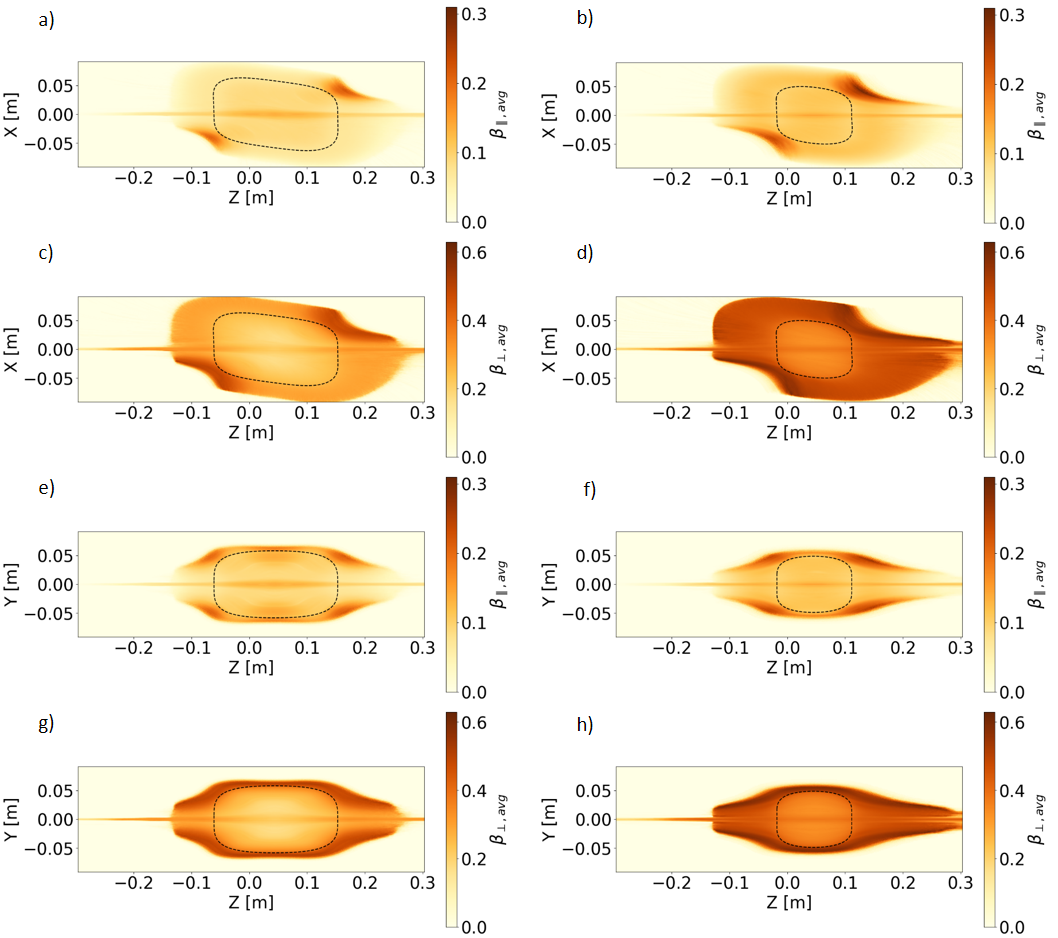}
\caption{Average velocity distribution and ECR zone (black dotted line) \textbf{a)} for parallel component in configuration $C_3$ and XZ plane at y = 0; \textbf{b)} for parallel component in configuration $C_8$ and XZ plane at y = 0; \textbf{c)} for perpendicular component in configuration $C_3$ and XZ plane at y = 0; \textbf{d)} for perpendicular component in configuration $C_8$ and XZ plane at y = 0; \textbf{e)} for parallel component in configuration $C_3$ and YZ plane at x = 0; \textbf{f)} for parallel component in configuration $C_8$ and YZ plane at x = 0; \textbf{g)} for perpendicular component in configuration $C_3$ and YZ plane at x = 0; \textbf{h)} for perpendicular component in configuration $C_8$ and YZ plane at x = 0.}
\label{fig:v_densityplot_long}
\end{figure*}

Eq.~\ref{eq:mub} shows that an ECRIS plasma with a high $B_{min}$ contains electrons with larger average $\beta_{\perp}$.  It is worth noting that the peripheral hump observed in all the curves of Fig.~\ref{fig:4_v}(b) for $r \geq r_{ecr}$ is located in regions where relativistic electrons (with a relativistic factor $\gamma>1$) can continue to resonantly gain energy on closed magnetic surfaces, where  $B=\gamma B_{ecr}$.
\\
Variations of $B_{min}$ from 0.3 T to 0.8 T (cases $C_3$, $C_6$ and $C_8$) strongly affect all the velocity profiles. 
A general increase of $\beta_{\perp}$ by 0.2 is observed all along the radial profile (Fig.~\ref{fig:4_v}(b)) when $B_{min}$ passes from 0.3 T to 0.8 T ($C_3$ to $C_8$).
The peripheral area with $r \geq r_{ecr}$, which is naturally enlarging when $B_{min}$ increases, contains the most energetic electrons. 
In order to assess more precisely the growth of the volume occupied by hot electrons with larger values of $B_{min}$, complementary spatial velocity distributions for $\beta_{||}$ and $\beta_{\perp}$ in the XZ plane (for y = 0) and in the YZ plane (for x = 0) are proposed in Fig.~\ref{fig:v_densityplot_long}. The left and right columns correspond respectively to the cases $C_3$ ($B_{min}$ = 0.3 T) and $C_8$ ($B_{min}$ = 0.8 T). In each plot, the ECR zone contour ($B_{ecr}\approx$ 1 T) is represented with a dashed black line. The XZ plane is oriented to contain an hexapole pole axis, while the YZ plane is located in between two poles.
The most important features to consider comparing case $C_3$ and $C_8$ are the following:
\begin{itemize}
 \item the size of the ECR volume;
 \item the general increase of $\beta_{\perp}$ inside the whole ECR volume;
 \item the comparable total volume where hot electrons are present in both configurations;
 \item the dramatic increase of the volume available, populated by energetic electrons (high $\beta_{\perp}$), around the ECR zone in the $C_8$ case with respect to $C_3$.
\end{itemize}

\begin{figure}[!ht]
\centering
\includegraphics[width=0.48\textwidth]{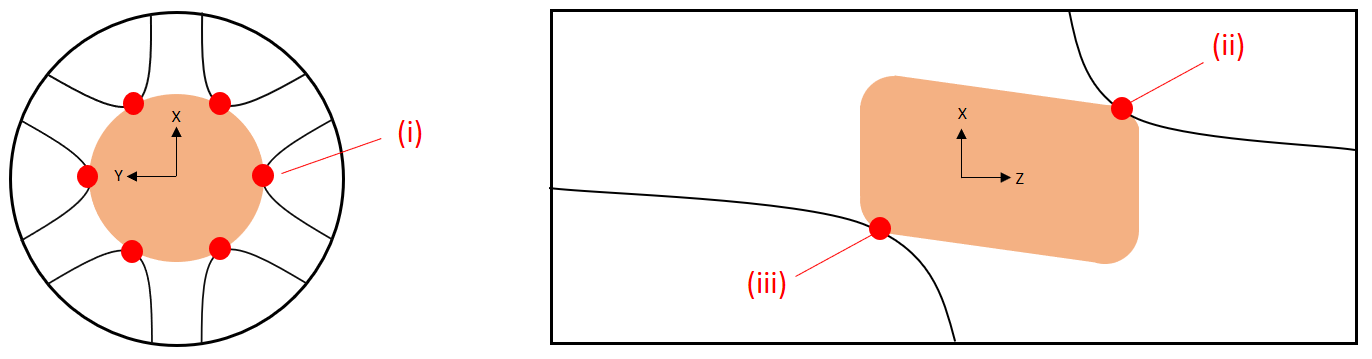}
\caption{Schematic representation of the locations where the magnetic field lines are tangent to the ECR surface. This behavior enhances ECR heating, as better explained in the text for regions (i), (ii) and (iii). }
\label{fig:ECRheatingspots}
\end{figure}

In ECRIS, it is observed that the hottest electrons are mostly accelerated and confined along the field lines tangent to the ECR zone. Electrons satisfying this condition are heated by the ECR mechanism on distances several times longer than the ones in regions where the field line crosses the ECR surface with a shallow magnetic field gradient.
The geometrical places where this tangency condition is met are the following (see Fig.~\ref{fig:ECRheatingspots}):

\begin{enumerate} [label=\roman*)]
    \item Six spots at $z=z_{min}$ and $r=r_{ecr}$, located at the azimuthal angles in between the hexapole poles (shifted by 30° from the poles). The electrons in these regions are trapped on short magnetic field lines, connecting one pole directly to the next one.
    \item Three spots on the extraction side of the plasma chamber, in the longitudinal planes containing even poles ($\theta=0\degree, \; 120\degree, \; 240\degree$) , close to the location where the axial resonance condition is satisfied ($B_z(z)=B_{ecr}$ and $r\approx r_{ecr}$). The electrons in these regions are trapped along very long field lines, connecting a hexapole pole at $r=r_W$ and the extraction axial plane at $z=z_{ext}$.
    \item Three spots on the injection side of the plasma chamber, in the longitudinal planes containing odd poles ($\theta=60\degree, \; 180\degree, \; 300\degree$).  The electrons in these regions are also trapped along extended field lines, connecting a hexapole pole at $r=r_W$ and the injection axial plane at $z=z_{inj}$. 
\end{enumerate}

The important magnetic field gradients along the field lines outside the ECR zone result in a strong electrons azimuthal drift: the hot electrons passing close to spots (i), (ii) and (iii) proceed to populate all the peripheral field lines around the ECR zone, and successively shift from one hexapole pole to the next. 
This explains the presence of high velocity particles surrounding the whole ECR zone in the YZ planes (Fig.~\ref{fig:v_densityplot_long}(e), (f), (g) and (h)), expanding toward both the extraction and injection direction.


\subsection{Electron Energy Spatial Distribution}
\label{sec:energy-spacial}

\begin{figure*}[!p]
\centering
\includegraphics[width=0.845\textwidth, valign=t]{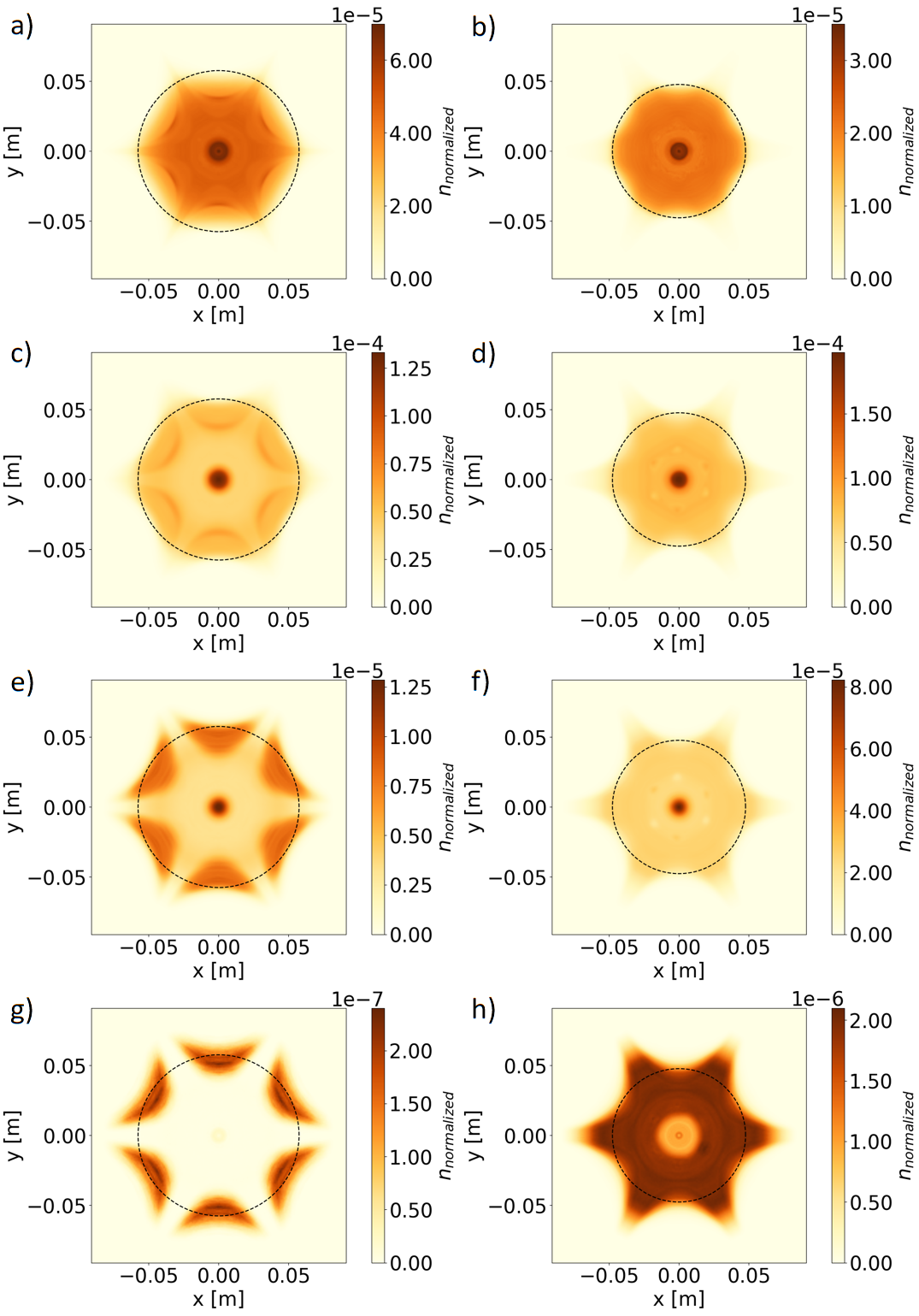}
\caption{Average electron density distribution and ECR zone (black dotted line) in the XY plane at $B(z_{min})=B_{min}$ split by energy group for the configurations $C_3$ (left) and $C_8$ (right) \textbf{a)} T $\leq$ 10 keV ($C_3$); \textbf{b)} T $\leq$ 10 keV ($C_8$); \textbf{c)} 10 keV $\leq$ T $\leq$ 100 keV ($C_3$); \textbf{d)} 10 keV $\leq$ T $\leq$ 100 keV ($C_8$); \textbf{e)} 100 keV $\leq$ T $\leq$ 300 keV ($C_3$); \textbf{f)} 100 keV $\leq$ T $\leq$ 300 keV ($C_8$); \textbf{g)} T $\geq$ 300 keV in configuration ($C_3$); \textbf{h)} T $\geq$ 300 keV ($C_8$). \textbf{N.B.} The normalization factor applies to the same configuration.}
\label{fig:n_densityplot_T}
\end{figure*}

In order to assess the previous result that a high $B_{min}$ favors the accumulation of hot electron population in the region surrounding the smaller ECR zone, the electron density distribution has been partitioned into 4 energy groups and plotted in Fig.~\ref{fig:n_densityplot_T} for the XY plane at $z=z_{min}$ ($B(z_{min})=B_{min}$) for both configurations $C_3$ ($B_{min}$ = 0.3 T, left column) and $C_8$ ($B_{min}$=0.8 T, right 
column).
The energy groups are the following: 
\begin{enumerate}
 \item $T\leq$ 10 keV (first row of Fig.~\ref{fig:n_densityplot_T});
 \item 10 $\leq T\leq $ 100 keV (second row of Fig.~\ref{fig:n_densityplot_T});
 \item 100 $\leq T\leq $ 300 keV (third row of Fig.~\ref{fig:n_densityplot_T});
 \item $T\geq$ 300 keV (fourth row of Fig.~\ref{fig:n_densityplot_T}).
\end{enumerate}
The ECR zone boundary is displayed with a black dashed line in each plot. While the spatial energy density distribution of electron is quite similar for the first two groups (hence for $T\leq$ 100 keV), the difference becomes significant at higher energy. For the case $C_3$ ($B_{min}$ = 0.3 T), the very hot electrons ($T\geq$ 300 keV) are concentrated close to the outer radius of the ECR zone and do not penetrate further into the plasma core.
On the other hand, the same electron group in case $C_8$ ($B_{min}$ = 0.8 T) is distributed on a very large area, covering almost the whole inner ECR volume, except for the axially confined electrons with $r \leq r_L$. Furthermore, the region occupied by the very hot electron population expands also in the volume surrounding the ECR zone, up to the radial wall.

Since the number of propagated electrons is the same in every configuration, it is possible to compare the estimated densities considering an error in the order of 20\%, assessed as the relative difference of total electron count for the same surface and integration time. Thus, the density of the very hot electron population ($T\geq$ 300 keV) in the case $C_8$ is estimated to be an order of magnitude higher than in the case $C_3$.


\section{Results for bremsstrahlung emission}


\subsection{Photons Count}

\begin{table}[b]
\caption{$N_{150}$ and $T_{150}$ are respectively the proportion of the electrons reaching 150 keV and the mean electron lifetime needed to reach such energy in the Monte Carlo simulation. $E_{150}$ is the ratio between the two aforementioned quantities, while $P_{150}$ is the number of photons generated per electron above 150 keV. Both $E_{150}$ and $P_{150}$ have been normalized with respect to the corresponding values of configuration $C_3$.}
\begin{ruledtabular}
\begin{tabular}{ccccccc}
Case & $B_{min}$ & $B_{ext}$ & $N_{150}$ & $T_{150}$& $E_{150}$ &$P_{150}$\\
&(T)&(T)& (\%)& ($\mu$s) & (a.u.)&(a.u.)\\
\colrule
$C_1$&0.3 &2.0 & 5.48 & 398 &0.99& 1.03\\
$C_2$&0.3 &2.5 & 5.70 & 400 &1.03& 0.98\\
$C_3$&0.3 &2.2 & 5.56 & 399 &1.00& 1.00 \\
$C_4$&0.4 &2.2& 6.19 & 369 &1.21& 2.27\\
$C_5$&0.5 &2.2& 5.95 & 309 &1.39& 2.84\\
$C_6$&0.6 &2.2& 5.44 & 247 &1.58& 2.69\\
$C_7$&0.7 &2.2& 4.47 & 193 &1.66& 3.09\\
$C_8$&0.8 &2.2& 5.40 & 156 &2.49& 4.72\\
$C_9$&0.9 &2.2& 7.43 & 138 &3.86& 5.36\\
\end{tabular}
\end{ruledtabular}
\label{tab:E150keVvsBmin}
\end{table}

Table~\ref{tab:E150keVvsBmin} presents the evolution of the ratio of electrons reaching 150 keV (noted $N_{150}$) for the 9 magnetic cases studied. One can note that $N_{150}$ remains almost constant for all the cases, in the range 4.5\%~-~7.5\%. On the other hand, the mean time for electrons to reach 150~keV, noted $T_{150}$, dramatically decreases when $B_{min}$ increases.
This trend is clearly related to the higher electron energies observed with growing values of $B_{min}$, as extensively discussed in Section~\ref{sec-simulation-Edistr}, \ref{sec-simulation-vdistr} and \ref{sec:energy-spacial}. Indeed, the similar electron temperatures in configurations with different $B_{ext}$ lead to a constant behavior for $T_{150}$ (profiles C1, C2 and C3).

\begin{figure}[b]
\includegraphics[width=\linewidth, valign=t]{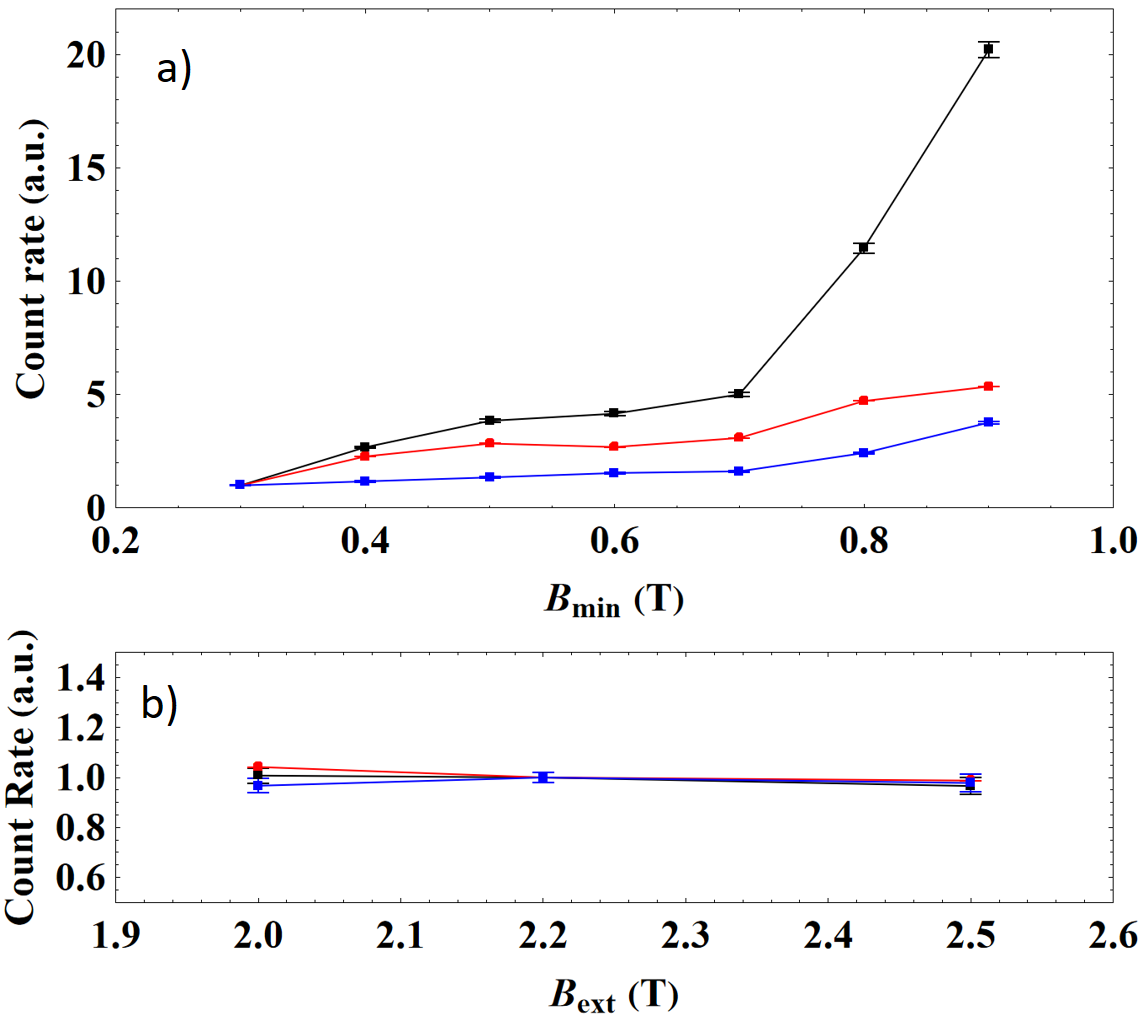}
\caption{Relative evolution of the photon count rate as a function of $B_{min}$ (a) and $B_{max} (b)$. Black curve: total photon count rate. Blue curve: Electron rate reaching 150 keV ($E_{150}$). Red Curve: photon production rate per electron above 150 keV ($P_{150}$). See text for details.}
\label{fig:photoncountvsBmin}
\end{figure}

Assuming that the plasma density is constant for all the magnetic cases investigated, the rate of electrons reaching an energy larger than 150 keV can be estimated as  
\begin{equation}
    E_{150}=\frac{N_{150}}{T_{150}}\;.
\end{equation}
The relative values of $E_{150}$, normalized to the case $C_3$, are available in Table~\ref{tab:E150keVvsBmin}. 
The  number of bremsstrahlung photons emitted per electron above 150~keV and normalized to the case $C_3$, noted $P_{150}$, have also been added to Table~\ref{tab:E150keVvsBmin}. The total photon flux above 150 keV is finally estimated as the product $E_{150} \cdot P_{150}$. The relative evolutions of $E_{150}$, $P_{150}$ and their product are plotted in blue, red and black respectively in Fig.~\ref{fig:photoncountvsBmin}, as a function of $B_{min}$ (a) and $B_{ext}$ (b).
In Fig.~\ref{fig:photoncountvsBmin}(a), it can be noticed a slow increase of the total photon count rate up to $B_{min}$ = 0.7 T, followed by a sharp increase for 0.8 T and 0.9 T, where the relative rate exceeds 20 with respect to $B_{min}$ = 0.3 T. No change in the photon count has been observed when $B_{ext}$ is varied (see Fig.~\ref{fig:photoncountvsBmin}(b)).

\subsection{Photons Directionality}

\begin{figure}[t]
\includegraphics[width=\linewidth, valign=t]{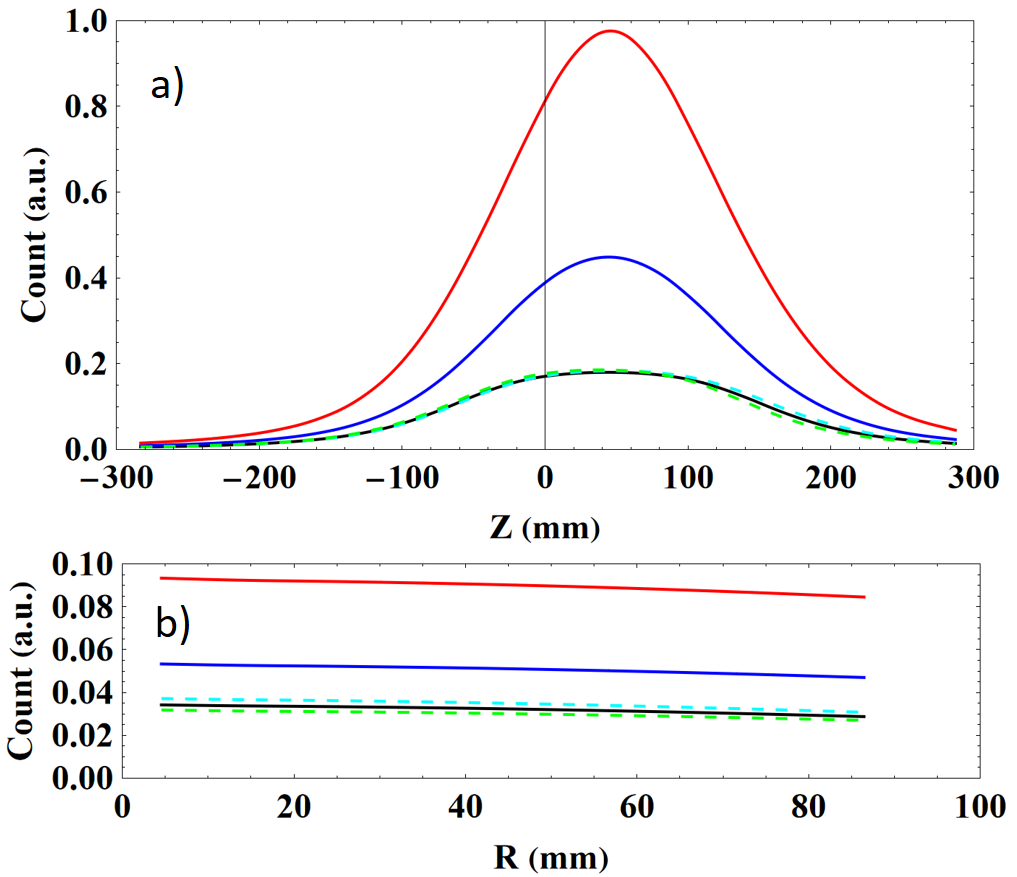}
\caption{Photon count distribution per unit surface (a) at the cavity radius $r$ = 91 mm as a function of the source axis $z$ and (b) as a function of the radius in the extraction plane ($z=z_{ext}$). Black line: profile $C_3$ ($B_{min}$ = 0.3 T, $B_{ext}$ = 2.2 T). Blue line: profile $C_6$ ($B_{min}$ = 0.6 T, $B_{ext}$ = 2.2 T). Red line: profile $C_9$ ($B_{min}$ = 0.9T, $B_{ext}$ = 2.2 T). Dashed cyan line: profile $C_1$ ($B_{min}$ = 0.3 T, $B_{ext}$ = 2.0 T). Dashed green line: profile $C_2$ ($B_{min}$ = 0.3 T, $B_{ext}$ = 2.5 T).}
\label{fig:photondensity}
\end{figure}

\begin{figure}[b]
\includegraphics[width=\linewidth, valign=t]{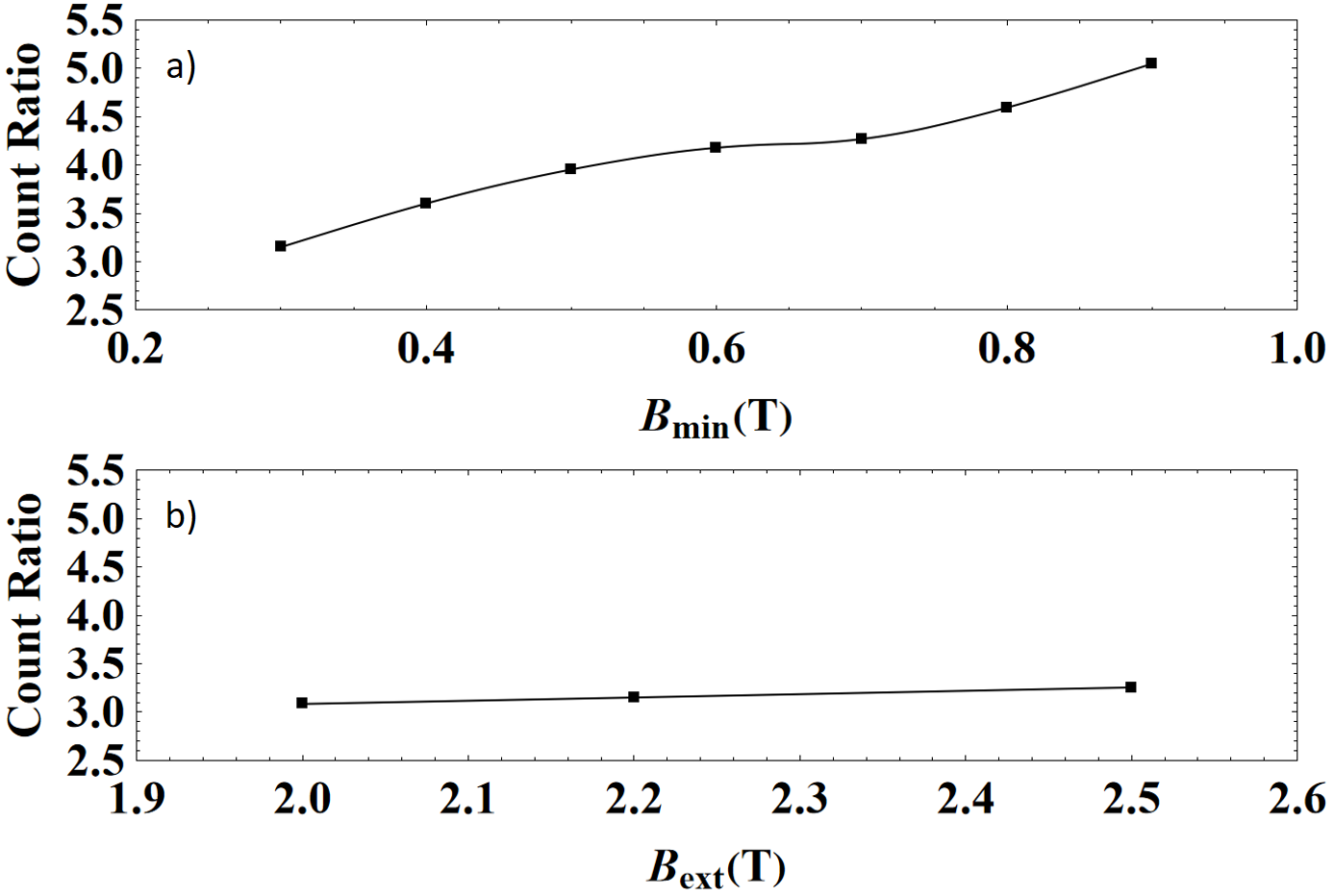}
\caption{Radial to axial photon count ratio per electron generated as a function of (a) $B_{min}$ and (b) $B_{ext}$.}
\label{fig:countphotonratio}
\end{figure}

The radial photon count distribution per unit surface, measured at the chamber wall (radius $r$ = 91 mm) as a function of the source axis direction $z$,  is displayed in Fig.~\ref{fig:photondensity}(a) for configuration $C_3$ ($B_{min}$ = 0.3 T, black line), $C_6$ ($B_{min}$ = 0.6 T, blue line) and $C_9$ ($B_{min}$ = 0.9 T, red line). The radial photon flux reaches a maximum around $z=z_{min}$, corresponding to the region where $B(z_{min})=B_{min}$.
The photon emission mainly extends within the axial range of the ECR zone, with a broader photon distribution for $B_{min}=0.3$ T and a narrower one for $B_{min}=0.9$ T.
The dashed cyan ($C_1$, $B_{ext}$ = 2.0 T) and green plots ($C_2$, $B_{ext}$ = 2.5 T) represent the effect due to variations of $B_{ext}$ with respect to the black line ($C_3$, $B_{ext}$ = 2.2 T). As expected, increasing $B_{ext}$ has an insignificant effect on the photon density profile.
\\
The axial photon count distribution per unit surface, taken at the extraction plane ($z=z_{ext}$), is presented in Fig.~\ref{fig:photondensity}(b) with the same color plot convention as Fig.~\ref{fig:photondensity}(a). In the latter, a quasi-flat photon density flux per unit surface is observed as a function of the radius, independently of $B_{min}$ and $B_{ext}$.
The axial photon count intensity increases with $B_{min}$, in the same proportion as the radial case. Once again, variations in $B_{ext}$ provide no significant effect on the photon count.
Fig.~\ref{fig:countphotonratio} presents the evolution of the radial to axial photon count ratio per unit surface, as a function of $B_{min}$ (Fig.~\ref{fig:countphotonratio}(a)) and $B_{ext}$ (Fig.~\ref{fig:countphotonratio}(b)). The photon count ratio increases from 3 to 5 when $B_{min}$ grows from 0.3 T to 0.9 T, while it remains approximately flat when only $B_{ext}$ is varied.

\subsection{Photon Energy Spectrum}

\begin{figure}[t]
\includegraphics[width=\linewidth, valign=t]{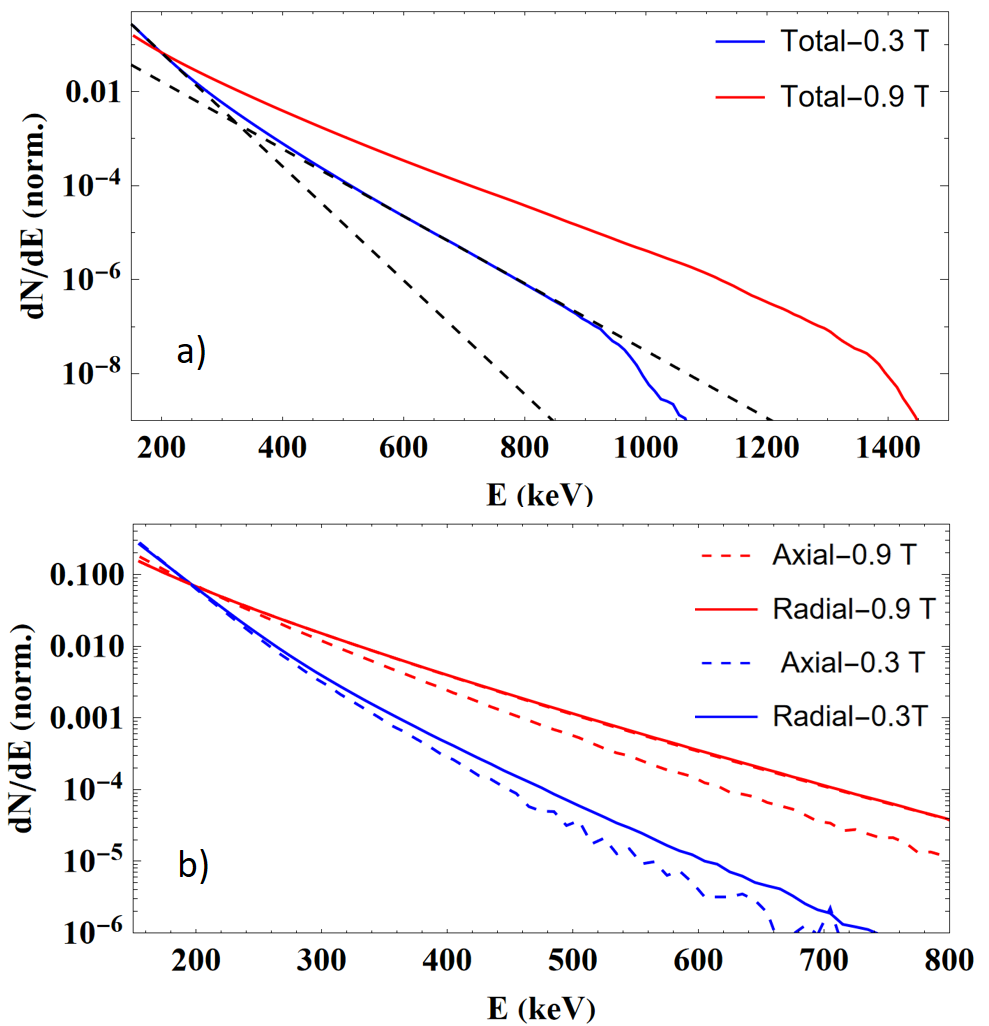}
\caption{Examples of normalized PEDF measured. (a) Total bremsstrahlung PEDF for $B_{min}$ = 0.3 T ($C_3$, blue curve) and $B_{min}$ = 0.9 T ($C_9$, red curve). The dashed lines are examples of local spectral temperature fits to measure $T_{S1}$ and $T_{S2}$. (b) Axial (dashed lines) and radial (solid lines) PEDF measured for $B_{min}$ = 0.3 T ($C_3$, blue curve) and $B_{min}$ = 0.9 T ($C_9$, red curve).}
\label{fig:dNdEphoton}
\end{figure}

Fig.~\ref{fig:dNdEphoton}(a) presents the total normalized bremsstrah\-lung photon energy distribution function (PEDF) for $B_{min}$ = 0.3 T ($C_3$) and 0.9 T ($C_9$). Unsurprisingly, the general PEDF obtained does not follow a steady exponential decay law $f(E)\propto e^{-E/kT_{S}}$, where $T_{S}$ is the spectral temperature. Such behavior relates to the EEDF discussed in Section~\ref{sec-simulation-Edistr}, which does not follow a Maxwell-Boltzmann distribution. As a consequence, the corresponding PEDF cannot be fitted with an exponential decay model. Therefore, the PEDF has been interpolated using two exponential decay curves, with spectral temperatures labeled $T_{S1}$ and $T_{S2}$ respectively.
$T_{S1}$ and $T_{S2}$ have been obtained by fitting photon data with energies 150 keV $\leq E \leq$ 500 keV and $E>$ 500 keV respectively.
\\
An example of the double exponential decay fit is displayed in Fig.~\ref{fig:dNdEphoton}(a) with dashed black lines. For energies above a limiting value, the simulated PEDF steeply drop and do not follow the fitted curve anymore. This behavior is understood as a statistical limitation rather than a physical one.
The generation of a larger number of electrons is expected to populate the PEDF up to higher maximum energies, with a much lower probability of occurrence. A higher photon statistics for the highest energy part of the spectrum is unreachable within a reasonable computational time of the MC simulation. Moreover, the PEDF reaches higher maximum photon energies for larger values of $B_{min}$ (see Fig.~\ref{fig:dNdEphoton}(a)). Since the number of generated electrons is the same for each magnetic configuration, it can already be deduced that the probability of generating more energetic photons (and hence electrons) increases with $B_{min}$.

\begin{table} [!t]
\caption{Results of spectral temperatures ($T_{S1}$ and $T_{S2}$) and photon count per impinging electron, measured as a function of $B_{min}$ and $B_{ext}$ for the total emitted PEDF and both the radial and axial telescopes.}
\begin{ruledtabular}
\begin{tabular}{lccccr}
\multicolumn{6}{c}{\textrm{Total photon emission}}\\
\textrm{Case} &\textrm{$B_{min}$} &\textrm{$B_{ext}$} & \textrm{$T_{S1}$} & \textrm{$T_{S2}$} &\textrm{Count} \\
& \textrm{[T]} & \textrm{[T]} & \textrm{[keV]} & \textrm{[keV]} &\textrm{/$e^-$} \\
\colrule
$C_1$&0.3&2.0 & $36.17\pm 0.09$ & $58.02\pm 0.26$ & 28866\\
$C_2$&0.3&2.5 & $35.28\pm 0.09$ & $54.02\pm 0.10$ & 27861\\
$C_3$&0.3 &2.2&$35.78\pm 0.10$ & $62.20\pm 0.09$ & 27986\\
$C_4$&0.4 &2.2&$48.57\pm 0.12$ & $62.00\pm 0.08$ & 70009\\
$C_5$&0.5 &2.2&$55.12\pm 0.28$ & $67.72\pm 0.17$ & 76850\\
$C_6$&0.6 &2.2&$54.31\pm 0.35$ & $78.00\pm 0.4$ & 57777\\
$C_7$&0.7 &2.2&$54.60\pm 0.32$ & $84.53\pm 0.23$ & 56654\\
$C_8$&0.8 &2.2&$58.83\pm 0.26$ & $87.92\pm 0.15$ & 117737\\
$C_9$&0.9 &2.2&$60.74\pm 0.28$ & $91.68\pm 0.10$ & 120407\\
\colrule
\noalign{\vskip 1.5mm}
\multicolumn{6}{c}{\textrm{Radial photon emission}}\\
Case &\textrm{$B_{min}$} &\textrm{$B_{ext}$} & \textrm{$T_{S1}$} & \textrm{$T_{S2}$} &\textrm{Count} \\
& \textrm{[T]} & \textrm{[T]} & \textrm{[keV]} & \textrm{[keV]} &\textrm{/$e^-$} \\
\colrule
$C_1$&0.3&2.0 & $33.03\pm 0.03$ & $57.64\pm 0.48$ & 125\\
$C_2$&0.3&2.5 & $32.16\pm 0.03$ & $53.63\pm 0.63$ & 129\\
$C_3$&0.3&2.2 & $32.26\pm 0.04$ & $59.50\pm 0.40$ & 121\\
$C_4$&0.4&2.2 & $48.46\pm 0.13$ & $55.79\pm 0.25$ & 427\\
$C_5$&0.5&2.2 & $55.34\pm 0.27$ & $67.90\pm 0.18$ & 508\\
$C_6$&0.6&2.2 & $54.71\pm 0.34$ & $78.60\pm 0.50$ & 394\\
$C_7$&0.7&2.2 & $55.15\pm 0.34$ & $83.41\pm 0.33$ & 370\\
$C_8$&0.8&2.2 & $59.62\pm 0.27$ & $87.14\pm 0.27$ & 828\\
$C_9$&0.9&2.2 & $60.73\pm 0.29$ & $89.66\pm 0.16$ & 935\\
\colrule
\noalign{\vskip 1.5mm}
\multicolumn{6}{c}{\textrm{Axial Photon Emission}} \\
\textrm{Case} &\textrm{$B_{min}$} &\textrm{$B_{ext}$} & \textrm{$T_{S1}$} & \textrm{$T_{S2}$} &\textrm{Count} \\
& \textrm{[T]} & \textrm{[T]} & \textrm{[keV]} & \textrm{[keV]} &\textrm{/$e^-$} \\
\colrule
$C_1$&0.3&2.0 & $30.84\pm 0.02$ & $44.50\pm 0.93$ & 3.6\\
$C_2$&0.3&2.5 & $30.52\pm 0.02$ & $47.20\pm 0.74$ & 3.6\\
$C_3$&0.3&2.2 & $30.43\pm 0.03$ & $48.20\pm 2.30$& 3.5 \\
$C_4$&0.4&2.2 & $44.36\pm 0.11$ & $47.20\pm 0.60$& 10.0\\
$C_5$&0.5&2.2 & $48.81\pm 0.23$ & $62.26\pm 0.23$& 9.9\\
$C_6$&0.6&2.2 & $46.52\pm 0.27$ & $76.50\pm 0.27$& 6.8\\
$C_7$&0.7&2.2 & $46.66\pm 0.24$ & $78.00\pm 0.70$& 6.2 \\
$C_8$&0.8&2.2 & $50.39\pm 0.18$ & $79.00\pm 0.70$& 12.9 \\
$C_9$&0.9&2.2 & $50.96\pm 0.21$ & $76.90\pm 0.70$ &13.2 \\
\end{tabular}
\end{ruledtabular}
\label{tab:Ts1Ts2vsCase}
\end{table}

\begin{figure}[t]
\includegraphics[width=\linewidth, valign=t]{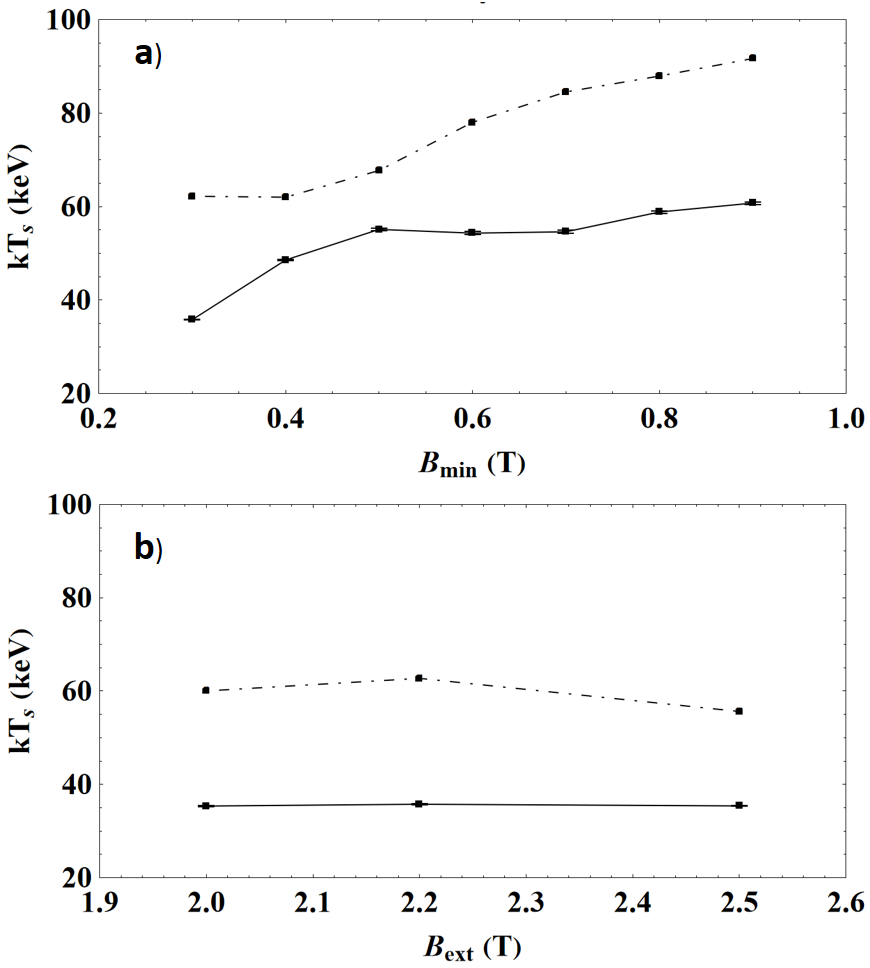}
\caption{Evolution of $kT_{S1}$ and $kT_{S2}$ as a function of (a) $B_{min}$ and (b) $B_{ext}$, evaluated from the total bremsstrahlung PEDF.}
\label{fig:kTs_vs_Bmin_Bext}
\end{figure}

A systematic analysis of the PEDF has been carried out for all the magnetic configurations. Fig.~\ref{fig:dNdEphoton}(b) presents the radial (solid line) and axial (dashed line) PEDF for the cases 0.3 T and 0.9 T, using the same color convention as Fig.~\ref{fig:dNdEphoton}(a). It is important to notice the lower statistics for these two telescopes, leading to a lower accuracy of the spectral temperatures fits. This consideration is especially true for the axial one, where only a few photons are detected in the high energy tail.
The complete results of the double temperature fit of the PEDF are listed in Table~\ref{tab:Ts1Ts2vsCase}, considering the total x-ray emission spectrum and both the radial and axial telescopes as a function of $B_{min}$ (cases $C_3$ to $C_9$) and $B_{ext}$ (cases $C_1$ to $C_3$). The spectral temperatures calculated with the total bremsstrahlung PEDF, more reliable thanks to the higher statistics, are plotted against $B_{min}$ and $B_{ext}$ in Fig.~\ref{fig:kTs_vs_Bmin_Bext}(a) and (b) respectively.

\begin{figure}[t]
\includegraphics[width=\linewidth, valign=t]{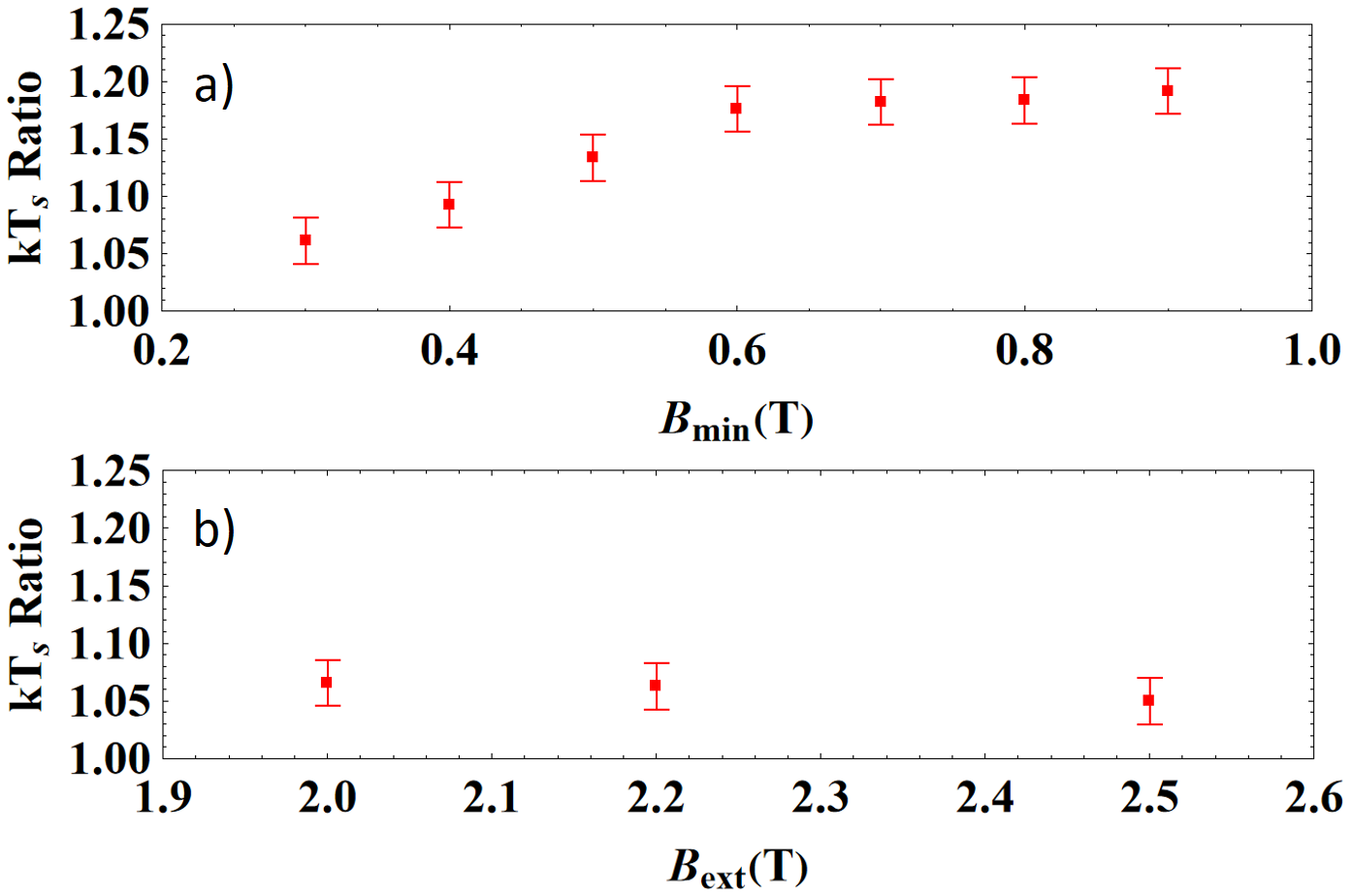}
\caption{Evolution of the radial to axial temperature ratio for $kT_{S1}$ as a function of $B_{min}$ (a) and $B_{ext}$ (b). }
\label{fig:kTs_Ratio}
\end{figure}

The higher temperature $kT_{S2}$ increases from 60 keV to 90 keV when $B_{min}$ grows from 0.3 T to 0.9 T. The associated trend for $kT_{S1}$ presents a flatter behavior, evolving from 30 keV to 60 keV. It is worth noting that the evolution of $kT_{S2}$ with $B_{min}$ is very similar to the experimental observations done in~\cite{Benitez17}, displaying a linear increase of the spectral temperature on the same range of $B_{min}$ from $\approx30$ keV to $\approx90$ keV. On the other hand, Fig.~\ref{fig:kTs_vs_Bmin_Bext}(b) exhibits a quasi-constant profile of both $kT_{S1}$ and $kT_{S2}$ with different values of $B_{ext}$.
Once again, this simulation result is in agreement with the experimental findings of~\cite{Benitez17}, which observed that $B_{min}$ is the only parameter influencing the  PEDF spectral temperature at high energy. It is also important to underline the fair consistency of the simulated photon spectral temperature with the electron temperatures of the confined electrons, plotted in Fig.~\ref{fig:kTe}.
The high photon statistics at 150~keV, the intense magnetic field of ECRIS and the presence of relativistic electrons in the plasma naturally lead to study the evolution of the ratio of the radial to axial temperature (for $kT_{S1}$, which presents the largest statistics), as a function of $B_{min}$ and $B_{ext}$ (see Fig.~\ref{fig:kTs_Ratio}(a) and (b) respectively). The ratio increases from 1.05 to 1.18 when $B_{min}$ passes from 0.3 to 0.6 T, reaching a plateau for higher $B_{min}$ values. On the other hand, no significant change on the temperature ratio is observed when $B_{ext}$ is varied.
The temperature ratio growth can be understood as a relativistic effect of the bremsstrahlung emission directionality, favoring forward emission toward the velocity vector when $\gamma>1$. Since the magnetic field in the source is mainly axial and the mirror trapped electrons have a high perpendicular velocity component, such temperature anisotropy is expected inside ECRIS.

\section{Conclusion}
\label{sec:conclusion}

The developed Monte Carlo code, dedicated to the study of the hot electron population in ECRIS, has successfully reproduced and refined former studies results achieved with more sophisticated self-consistent codes, including plasma ion dynamics, plasma potential and/or electromagnetic cavity mode~\cite{Mironov_2021,Mascali_3Dsymm}. This agreement is a consequence of the properties of the hot electron population simulated, which is weakly influenced by the plasma potential in the order of a few tens of volts. Moreover, the reproduction of results with a simple RF model suggests that the actual electron density in ECRIS plasma  is of the order of a few tens of \% of the cut-off density, confirming the hypothesis chosen for the study.
The unique high statistics achieved with the present Monte Carlo code brings extra refinement and new details on ECRIS studies, which were absent in the former published self-consistent studies featuring a much lower statistics.
\\
The bremsstrahlung emission inside an ECRIS is simulated for the first time. The general photon energy spectrum is well reproduced by a double exponential decay model. Such spectral shape has been reported in earlier experiments~\cite{Lamoureux_eibrem}. The simulated spectral temperature dependence of the high energy PEDF with $B_{min}$ successfully reproduces experimental measurements~\cite{Benitez17}. 
Additionally, the change of the extraction magnetic field $B_{ext}$ does not result in variations of the bremsstrahlung temperature, as already observed in experiments. Due to the symmetrical structure of the magnetic field, a change of $B_{inj}$ is not expected to modify the photon spectral temperature either, and thus has not been considered in the current work.
A higher radial spectral temperature by $\sim$~20\% with respect to the axial one is observed when $B_{min}$ exceeds 0.6 T. Moreover, the photon flux per unit surface is also enhanced by a factor of $\sim$ 20 when $B_{min}$ increases from 0.3 T to 0.9 T. This growth can be understood as a consequence of three phenomena:

\begin{enumerate}
    \item the larger bremsstrahlung emission yield in relativistic regime;
    \item the shorter time needed by electrons to reach relativistic energy through ECR heating, due to the shallower gradient around the ECR surface;
    \item the reduction of electron confinement time, which dramatically enhances the hot electron flux for a sustained plasma density.
\end{enumerate}

The increase of the emitted photon rate with $B_{min}$ has been indirectly measured experimentally through the parasitic heating of the superconducting magnet cold mass in ECRIS~\cite{leitner08}. The build up of a very hot energy tail with increasing $B_{min}$, responsible for the larger bremsstrahlung spectral temperature, can be explained by this MC investigation as a conjunction of the following factors:

\begin{enumerate}
    \item The decrease of the ECR zone radius when $B_{min}$ grows, allowing for a larger volume with $r_{ecr}\leq r \leq r_W$. In these regions, relativistic electrons can continue resonating on ECR surfaces (defined by the intensity $\gamma\cdot B_{ecr}$) for a long time before impinging the radial walls.
    \item The concomitant presence of long field lines traversing the (relativistic and non relativistic) ECR surfaces with shallow magnetic gradient, leading to enhanced electron heating and a larger electron accumulation.
    \item The general increase of $\beta_{\perp}$ inside the ECR plasma, and more specifically inside the ECR surface, where the growth is linked to the quasi-constancy of the averaged electron magnetic moment.
\end{enumerate}

It is remarkable that, despite the reduction of the electron confinement time with $B_{min}$, the electrons are reaching much higher kinetic energies. In other words, the characteristic ECR heating time reduces much faster than the confinement time when $B_{min}$ is increased, due to the presence of shallower gradients and a reduced ECR zone size.
All these behaviors are an indirect consequence of the magnetic properties of the minimum-B structure, composed by the axial magnetic field (defining $B_{min}$) and a radial multipole field, whose intensity increases with the square of the chamber radius.
Moreover, the constant behavior of the $\beta$ profiles when $B_{ext}$ is varied can be understood as a consequence of two factors:

\begin{enumerate}
    \item the unchanged geometry of the ECR zone, which is indeed mostly influenced  by $B_{min}$;
    \item the unchanged minimum magnetic mirror ratio $R$.
\end{enumerate}

Finally, this work shows that the diameter of the dense electron population,  centered around the ion source axis, scales with the plasma chamber diameter, provided that the ECRIS magnetic field confinement peak intensities follow the geometrical scaling. This observation corroborates the  choice to design a larger plasma chamber in the future ASTERICS ion source, in order to produce more intense ion beams.

\section*{ACKNOWLEDGMENTS}

This work is supported by Agence Nationale de la Recherche with the contract \# 21-ESRE-0018 EQUIPEX+ NEWGAIN.

\bibliography{Bibliography}

\cleardoublepage

\section*{Appendix}

\begin{figure}[t]
    \includegraphics[width=0.48\textwidth]{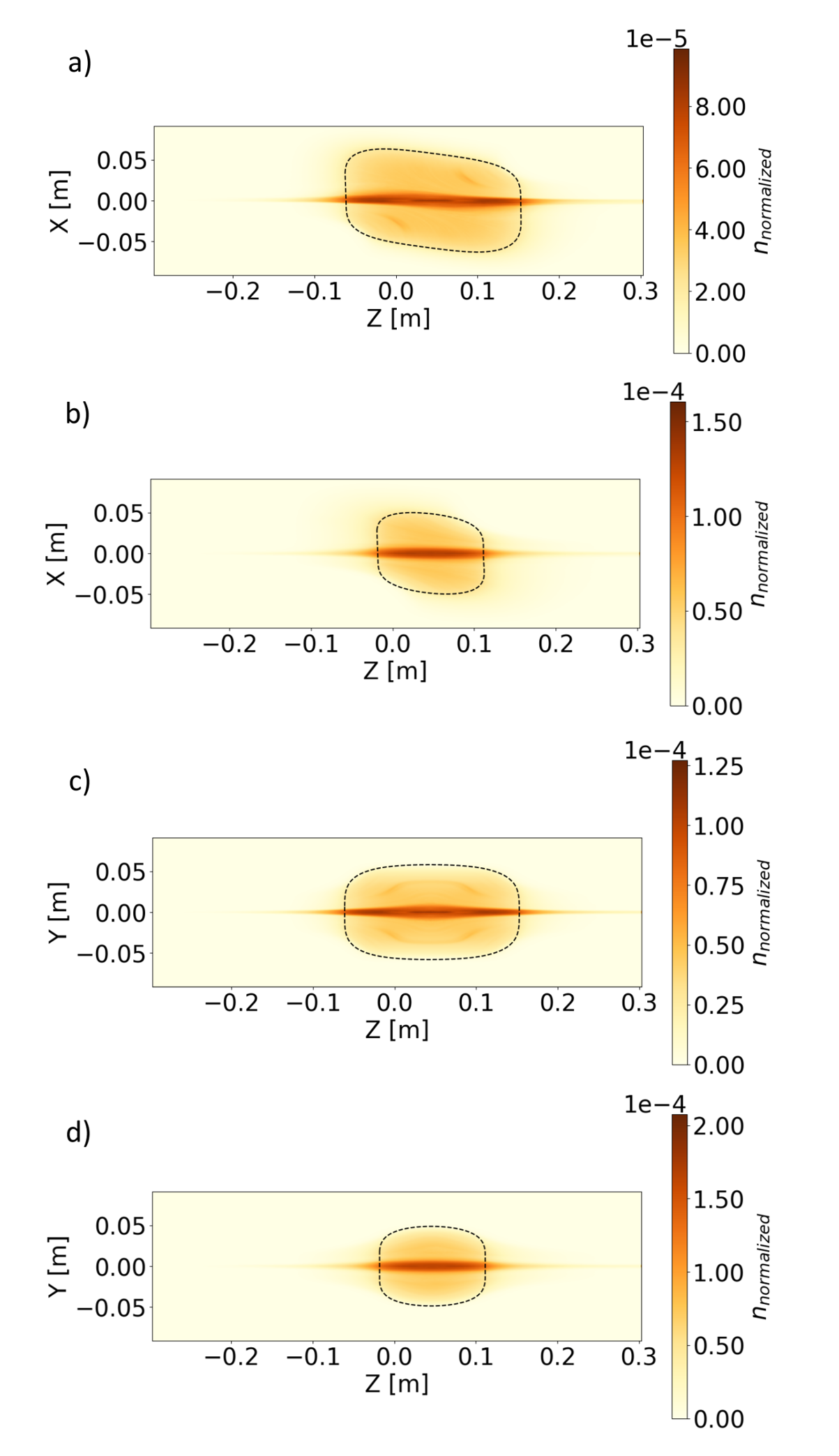}
    \caption{Average density distribution and ECR zone (black dotted line) \textbf{a)} in configuration $C_3$ and XZ plane at y = 0; \textbf{b)} in configuration $C8$ and XZ plane at y = 0; \textbf{c)} in configuration $C_3$ and YZ plane at x = 0; \textbf{d)} in configuration $C_8$ and YZ plane at x = 0.}
    \label{fig:n_densityplot_long}
\end{figure}

The appendix contains one table and one figure providing further details to the work. Table~\ref{tab:distr_position} presents the relative repartition of the particles when they either hit a wall or the program times out. This data is presented in Fig.~\ref{fig:conftime} and details are given in Section~\ref{sec:loss}. Fig.~\ref{fig:n_densityplot_long} presents two dimension density plots taken in the planes XZ and YZ for the magnetic cases C3 and C8. It can be noticed the presence of a dense electron population around the ion source axis in both cases.

\begin{table}[h!]
    \caption{\label{tab:distr_position}%
    Distribution of the final position of the electrons for the various magnetic axial profiles.
    The subscripts \textit{inj}, \textit{ext} and \textit{rad} refer respectively to the particles deconfined at the injection, extraction and radial walls, while \textit{conf} refers to the electrons which are still confined at the end of the simulation.
    }
    \begin{ruledtabular}
    \begin{tabular}{ccccc}
    Axial profile&
    \%$_{inj}$&
    \%$_{ext}$&
    \%$_{rad}$&
    \%$_{conf}$\\
    \colrule
    $C_1$ & 2.9 & 32.6 & 60.9 & 3.5\\
    $C_2$ & 3.2 & 17.3 & 76.0 & 3.6\\
    $C_3$ & 3.5 & 11.2 & 81.7 & 3.6\\
    $C_6$ & 4.5 & 27.8 & 66.5 & 1.3\\
    $C_8$ & 5.8 & 41.5 & 52.1 & 0.6\\
    \end{tabular}
    \end{ruledtabular}
\end{table}

\newpage

\clearpage

\end{document}